\documentclass[conference]{IEEEtran}
\IEEEoverridecommandlockouts
\usepackage{subcaption}
\usepackage{cite}
\usepackage{amsmath,amssymb,amsfonts}
\usepackage{algorithmic}
\usepackage{graphicx}
\usepackage{textcomp}
\usepackage{xcolor}
\def\BibTeX{{\rm B\kern-.05em{\sc i\kern-.025em b}\kern-.08em
    T\kern-.1667em\lower.7ex\hbox{E}\kern-.125emX}}
\begin{document}

\title{A Serverless Cloud-Fog Platform for DNN-Based Video Analytics with Incremental Learning
}


\author{
    \IEEEauthorblockN{Huaizheng Zhang\IEEEauthorrefmark{1}, Meng Shen\IEEEauthorrefmark{1}, Yizheng Huang\IEEEauthorrefmark{1}, Yonggang Wen\IEEEauthorrefmark{1}, Yong Luo\IEEEauthorrefmark{2}, Guanyu Gao\IEEEauthorrefmark{3}, Kyle Guan\IEEEauthorrefmark{4}}
    \IEEEauthorblockA{\IEEEauthorrefmark{1}Nanyang Technological University
    \\\IEEEauthorrefmark{1}\{huaizhen001, meng005, yizheng.huang, ygwen\}@ntu.edu.sg}
    \IEEEauthorblockA{\IEEEauthorrefmark{2}Wuhan University, \IEEEauthorrefmark{3}Nanjing University of Science and Technology, \IEEEauthorrefmark{4}K\&C Technologies Solutions
    \\\IEEEauthorrefmark{2}yong.luo@whu.edu.cn, \IEEEauthorrefmark{3}gygao@njust.edu.cn, \IEEEauthorrefmark{4}kcguan@gmail.com}
}
\maketitle

\begin{abstract}
Deep neural networks (DNN) based video analytics have empowered many new applications, such as automated retail and smart city. Meanwhile, the proliferation of fog computing systems provides system developers with more design options to improve performance and save cost. To the best of our knowledge, this paper presents the first serverless system that takes full advantage of the client-fog-cloud synergy to better serve the DNN-based video analytics. Specifically, the system aims to achieve two goals: 1) Provide the optimal analytics results under the constraints of lower bandwidth usage and shorter round-trip time (RTT) by judiciously managing the computational and bandwidth resources deployed in the client, fog, and cloud environment. 2) Free developers from tedious administration and operation tasks, including DNN deployment, cloud and fog's resource management. To this end, we design and implement a holistic cloud-fog system referred to as VPaaS (Video-Platform-as-a-Service) to execute inference related tasks. The proposed system adopts serverless computing to enable developers to build a video analytics pipeline by simply programming a set of functions (e.g., encoding and decoding, and model inference). These functions are then orchestrated to process video streaming through carefully designed modules. To save bandwidth and reduce RTT at the same time, VPaaS provides a new video streaming protocol that only sends low-quality video to the cloud. The state-of-the-art hardware accelerators and high-performing DNNs deployed at the cloud can identify regions of video frames that need further processing at the fog ends. At the fog ends, misidentified labels in these regions can be corrected using a light-weight DNN model. To address the data drift issues, we incorporate limited human feedback into the system to verify the results and adopt incremental machine learning to improve our system continuously. The evaluation of our system with extensive experiments on standard video datasets demonstrates that VPaaS is superior to several state-of-the-art systems: it maintains high accuracy while reducing bandwidth usage by up to 21\%, RTT by up to 62.5\%, and cloud monetary cost by up to 50\%. We plan to release VPaaS as open-source software to facilitate the research and development of video analytics.
\end{abstract}


\section{Introduction}


We are witnessing an unprecedented increase in high-resolution video camera deployment. These cameras, equipped with video analytics applications, continually collect high-quality video data and generate valuable insights. Many industries, such as automated retail, manufacturing, and smart city, rely on these video analytics to improve service efficiency and reduce operational expenditure \cite{emmons2019cracking, ananthanarayanan2019video}. For instance, thousands of cameras in a city combined with a traffic monitoring application are able to provide drivers with optimized routes to reduce traffic congestion \cite{bas2007automatic}. The key to such success lies in recent advances in deep neural networks (DNNs) that allow these applications to analyze video content with extremely high accuracy.


Utilizing DNN to analyze video content is not without its drawback. The DNN models (e.g., FasterRCNN101 \cite{ren2015faster}) that provide the highest accuracy can consist of hundreds of layers and millions of weight parameters. These extremely computationally intensive tasks thus rely on state-of-the-art hardware accelerators (e.g., GPUs or TPUs). Since client/edge devices are limited in their computational resources, the analytics tasks are executed in clouds equipped with the newest computational hardware so as to provide real-time feedback. As a consequence, the current video analytics pipeline needs first to stream videos to clouds, resulting in \textit{high bandwidth usage}. Moreover, many works \cite{ogden2019characterizing, zhang2020no} show that the video transmission time accounts for nearly half of the end-to-end processing time (the transmission time would increase even more in the presence of communication link outage or network congestion). As a result, users experience from \textit{long round-trip time (RTT)} from time to time. 


In order to address the aforementioned issues, many efforts have been invested in designing efficient video analytics systems \cite{chen2015glimpse, ran2018deepdecision, hung2018videoedge, zhang2018awstream, du2020server}. In general, these systems fall into two categories, the client-driven, and the cloud-driven methods, respectively. The client-driven method runs small models \cite{kang2017noscope, canel2019scaling} or simple frame differencing algorithms \cite{chen2015glimpse} on resource-limited client devices to filter video frames, only sending regions potentially having target objects to clouds for further processing. These methods suffer from missing important regions or sending redundant information due to the simple techniques used \cite{du2020server}. To overcome this issue, current systems \cite{zhang2017live, zhang2018awstream, du2020server} focus on utilizing cloud-driven methods that offload more model computational tasks to the cloud-side. For instance, CloudSeg \cite{wang2019bridging} sends low-resolution data from the client to the cloud, which runs a super-resolution model \cite{ahn2018fast} to recover a frame to high resolution for prediction. DDS \cite{du2020server} and SimpleProto \cite{pakha2018reinventing} design a multiple-round transmission method to save bandwidth while maintaining high accuracy.

\begin{figure}[]
    \centering
    \includegraphics[width=0.9\linewidth]{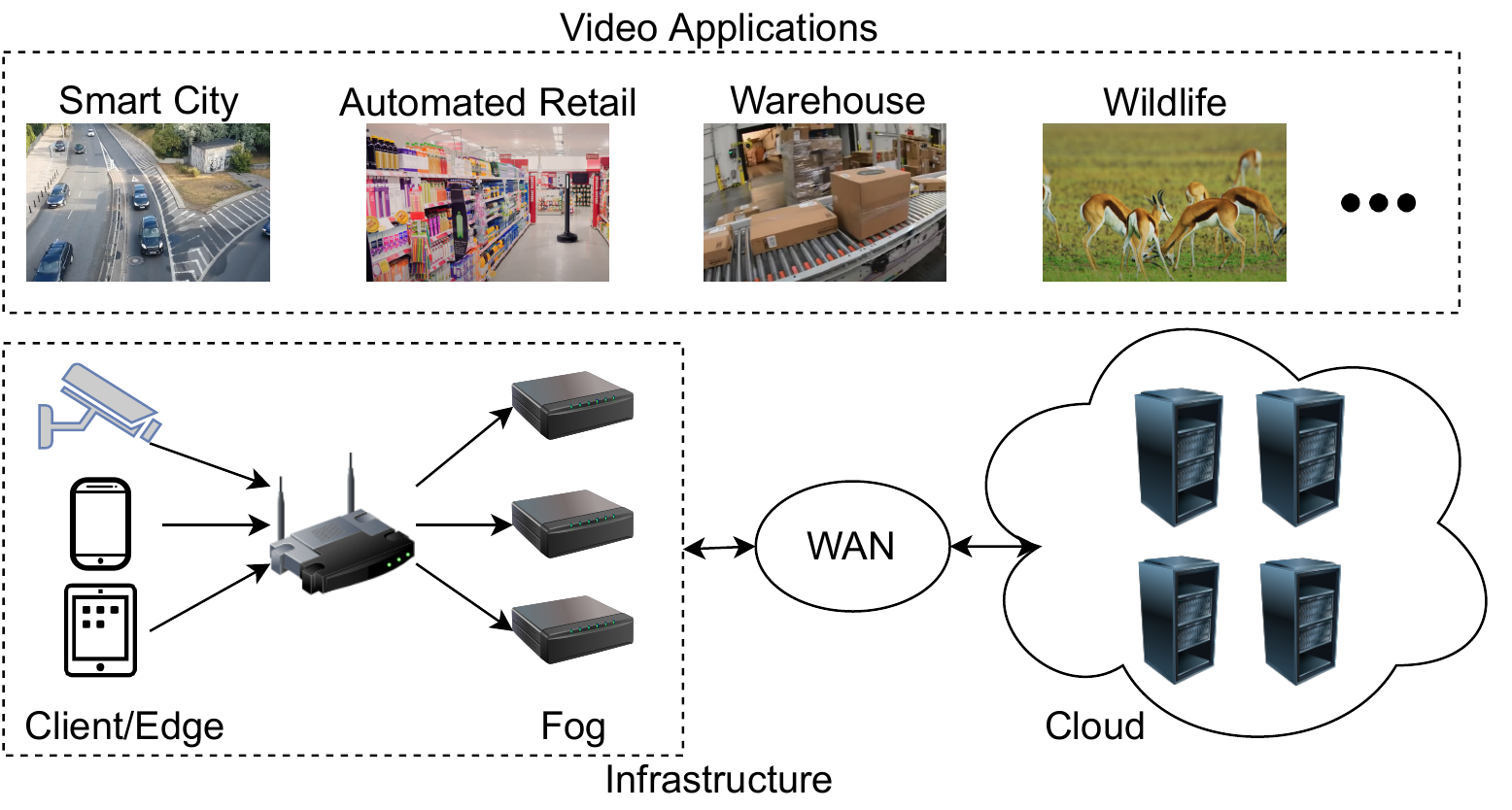}
    \caption{The client-fog-cloud infrastructure is widely deployed to support many video applications. In general, the clients equipped with high-resolution cameras have very limited computation resources and can not support DNNs' inference and video encoding and decoding very efficiently. The co-located fog nodes can only support lightweight DNN models; while the cloud can support computationally heavy DNN models.}
    \label{fig:real_system}
\end{figure}

Though these cloud-driven methods can potentially alleviate the bandwidth consumption to a certain degree, they still have drawbacks and leave much room to be improved. First, by trading bandwidth with cloud computational resources, these methods inadvertently increase cloud infrastructure and operational costs. For instance, DDS \cite{du2020server} runs multiple rounds of inference on a single frame to attain high accuracy, and CloudSeg \cite{wang2019bridging} needs extra cloud resources to run a super-resolution model to recover low-resolution frames. Moreover, these designs often incur higher latency overhead. Thus the performance of RTT can deteriorate in the presence of network congestion or unavailability of cloud resources. Second, most of these works over-rely on the performances of the state-of-the-art models (e.g., FasterRCNN101) to guide their design options \cite{jiang2018chameleon}. Specially, these single-stage models are often designed and trained to simultaneously perform several tasks (such as detection and classification) in an end-to-end fashion. These works overlook the fact that a carefully designed and trained multi-stage model that consists of models, each optimized only for a single task, can also achieve similar performance with less computational as well as bandwidth resources. Besides, these fixed and pre-trained models can easily meet the data drift issue when the training dataset distribution is different from that of online data. Third, deploying and operating these DNN-based video analytics pipeline are still non-trivial and require immense manual efforts. Even for a simple object detection task, developers have to undergo tedious administration and operation tasks such as resource management, DNN deployment, and so on.


We aim to design a cloud-driven platform for DNN-based video analytics with the following design guidelines to narrow these gaps. First, we utilize the widely deployed fog nodes as shown in Figure \ref{fig:real_system}, to minimize the cloud infrastructure cost and bandwidth usage while still achieving an accuracy performance comparable to those of the cloud-driven methods. The key insight is that a well designed and trained detection model can provide a very accurate target location even for a low-quality frame (though this model cannot classify the object). With object locations, we only need to run classification models, which require much fewer computation resources compared to detection models. Moving DNN inference tasks to fog nodes that are close to users can also reduce the transmission time. Second, involving a very limited human feedback to the system rather than relying exclusively on the best cloud models (namely the golden configuration in \cite{jiang2018chameleon}), the system should have the ability to evolve itself continuously to address the data drift issue and maintain the accuracy performance. Third, the often manual and mundane pipeline administration and resource management work should be kept minimal and even be avoided. The system should provide configurability and flexibility for users to easily build their video analytics pipelines and automate the operational tasks.


Following these guidelines and insights, we propose and develop VPaaS (Video-Platform-as-a-Service), a cloud-fog co-serving platform for DNN-based video analytics that takes the full benefits of the client-fog-cloud synergy. First, a client equipped with cameras sends high-quality videos to fog, where the videos will be re-encoded into low-quality videos. The low-quality videos are then sent to the cloud, where a state-of-art detection model is employed to analyze video content. The regions (within a video frame) with high confidence classification scores are considered as successfully identified. The coordinates of these regions with high confidence location scores are sent back to the fog node, where a lightweight classification model is employed to recognize these regions using high-quality frames. Second, to correct the miss-labeling caused by the best cloud model and continuously improve the system, we design an interface and data collector to collect human feedback and employ incremental machine learning to update fog models. Third, the whole system adopts the serverless computing design philosophy. We provide a set of functions to ease the development and deployment of a video analytics pipeline. Users can easily register and run their newly designed models and scheduling policies in our system that makes it easy to orchestrate both cloud and fog resources.

We deploy VPaaS into a testbed that emulates various real-world scenarios and evaluate system performance using multiple video datasets. Across these datasets, VPaaS achieves a comparable or higher accuracy while reducing the bandwidth usage by up to 21\%, RTT time by up to 62.5\%, and cloud cost by 50\%. Meanwhile, it can improve itself with very low resource consumption and update models with almost negligible overhead and bandwidth usage. We also conduct many case studies to show the ease of use, fault-tolerance, and scalability of our system. The contributions of this paper can be summarized as follows:
\begin{itemize}
    \item We develop a completed video processing platform termed VPaaS and conduct extensive experiments to verify its effectiveness.
    \item We implement a novel video streaming protocol that can save bandwidth usage while maintaining high accuracy by utilizing both cloud and fog resources.
    \item We employ incremental machine learning to improve system performance with minimal human efforts.
    \item To the best of our knowledge, VPaaS is the first serverless cloud-fog platform that provides a set of APIs to ease the development and deployment of video analytics applications.
\end{itemize}

The reminder of the paper is organized as follows. Section \ref{sec:background_motivation} introduces the background and related work. Section \ref{sec:system_design} provides the overview of VPaaS. Section \ref{sec:fog_coordinator} and Section \ref{sec:human_in_the_loop} detail the protocol and incremental learning design. Section \ref{sec:experiment} presents the evaluation results. Section \ref{sec:conclusion} summarizes the paper and discusses the future work.

\section{Background and Related Work}
\label{sec:background_motivation}

This section first introduces the DNN-based video analytics pipeline and our deployment scenarios and then surveys the related work both video analytics and serverless computing. 

\subsection{DNN-based Video Analytics}
\label{sec:dnn-video-analytics}

\begin{figure}
\begin{subfigure}{1.0\columnwidth}
  \centering
  \includegraphics[width=0.9\linewidth]{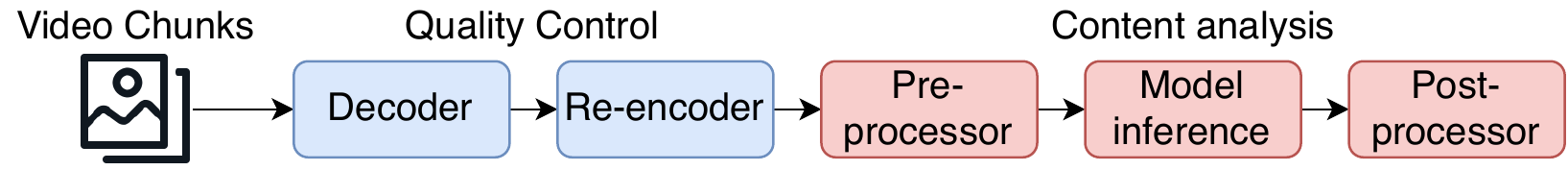}
  \caption{Vanilla Pipeline}
  \label{fig:general_pipeline}
\end{subfigure}
\begin{subfigure}{1.0\columnwidth}
  \centering
  \includegraphics[width=0.9\linewidth]{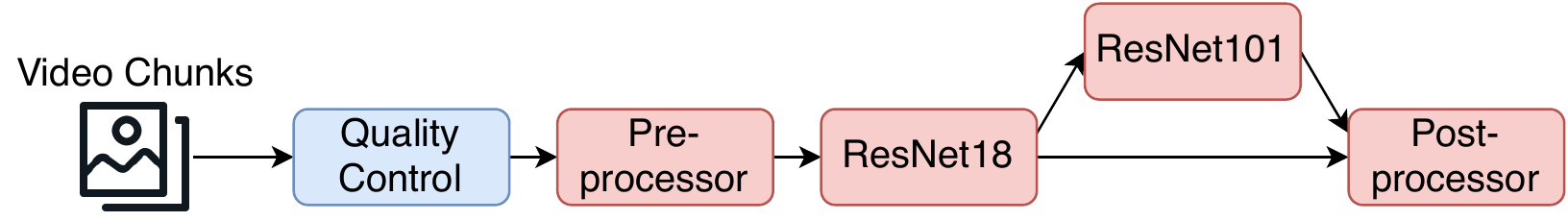}
  \caption{Cascade Pipeline}
  \label{fig:cascade_pipeline}
\end{subfigure}
\caption{The example video analysis pipelines. In general, a pipeline consists of two stages, quality control and content analysis. A video stream is re-encoded for bandwidth efficiency. Then it will be analyzed through three functions, pre-processing (e.g., key-frame extraction and resize), model inference (e.g., object detection), and post-processing (e.g., add bounding boxes).}
\label{fig:example_pipelines}
\end{figure}

\textbf{Video analytic pipeline.} Typical video analytics pipelines are shown in Figure \ref{fig:example_pipelines}. It contains two main stages, quality control and content analytics. In the quality control stage, a video chunk's quality setting will be adjusted via a decoder and encoder so that it can be transmitted in a bandwidth-efficient manner. Many settings, such as resolution, frame rate and quantization parameter (QP), can be chosen to improve or degrade video's quality \cite{zhang2018awstream,jiang2018chameleon,du2020server}.

Once the server (either in cloud or fog) receives a video chunk, it will start the content analytics process that consists of three steps, pre-processing, model inference, and post-processing.  The pre-processing step contains many functions, such as decoding and image manipulation. These functions decode a video chunk into many image frames and resize them so that they can be fed into DNN models for further processing. Many DNN models, such as object detection, object tracking, and image classification, can be employed to analyze video content. Each type of models has variants, each of which has different processing speeds and prediction accuracy. Users need to evaluate the performance - processing speed trade-off before deploying them to a device. 

\textbf{Video applications under consideration.} This paper considers quasi-real-time video applications such as automated retail video analytics, warehouse management, etc. Unlike the video analytics for self-driving, which have very stringent latency constraints, these video applications provide near real-time and high-quality feedback so the downstream tasks can be finished in time. For instance, in an automated retail store, the video monitoring system must process videos efficiently and accurately so the clearing system can help customers check out in time. Also, for modern warehouses (such as Amazon warehouses), this kind of video applications is crucial for efficient management of workflows.

\textbf{Client-Fog-Cloud infrastructure.} Figure \ref{fig:real_system} shows our system's deployment scenario, representing real-world applications. In this scenario, a client device is used to collect information and generate video chunks, which will be sent to backend devices for content analytics. The backend devices include fog nodes and cloud servers. The fog nodes are often directly connected to the end devices and have limited computational resources. In comparison, the cloud servers are located in data centers connected to edge devices and fog nodes via local-area and wide-area networks and are equipped with the high-performing hardware accelerators (e.g., NVIDIA V100 GPU) to perform video analytics with the state-of-the-art DNN models.

\subsection{Related Work}

\textbf{System for video analytics.} Many systems \cite{chen2015glimpse, ran2018deepdecision, hung2018videoedge, zhang2018awstream, jiang2018chameleon, wang2019bridging, du2020server} have been implemented to optimize the video analytics process with metrics such as latency, bandwidth, etc. They can be categorized into client/edge-driven methods and cloud-driven methods. Representative client/edge approaches include NoScope \cite{kang2017noscope}, FileterForward \cite{canel2019scaling}, etc. They use lightweight and customized DNN models to improve the processing throughput with a lower inference time. Glimpse \cite{chen2015glimpse} designs a frame difference analyzer to filter out frames and run a tracking model on the client for object detection.

In comparison, cloud-driven methods balance the trade-off between resources and accuracy to optimize systems. CloudSeg \cite{wang2019bridging} leverages a super-resolution model to improve the video quality for a high-accurate recognition on the cloud. VideoStorm \cite{zhang2017live} examines the impact of many system parameters on the final results and thus orchestrates GPU cluster resources more efficiently. AWSstream \cite{zhang2018awstream} aims to wide-area video streaming analytics with efficient profiler and runtime. DDS \cite{du2020server} designs a multiple-round video streaming protocol. However, these systems (1) ignore the widely deployed fog nodes to further optimize the system while VPaaS leverages the fog nodes with a new protocol design and (2) do not consider the data drift issue while VPaaS employs human-in-the-loop machine learning to improve models continuously. Besides, VPaaS is the first integrated serverless cloud-fog platform that provides many functions for usability, scalability, fault-tolerance, etc.

\textbf{Serverless Computing.} Serverless cloud computing is designed to automatically handle all system administration operations, including auto-scaling, monitoring, and virtual machine (VM) configuration, thus greatly simplifying cloud service development and deployment. Many systems have been proposed to leverage their properties for large-scale video processing. For instance, ExCamera \cite{fouladi2017encoding} utilizes AWS Lambdas for encoding massive videos in a few minutes. Sprocket \cite{ao2018sprocket} is a video processing (including both video encoding and content analytics) framework built on an AWS serverless platform. However, none of these systems include offloading computation to fog nodes, especially regarding DNN-based video analytics applications.

\section{System Design}
\label{sec:system_design}

This section first details the system design and workflow and then introduces the core components for video analytics pipeline with serverless computing.

\subsection{Design Goals}

\textbf{Seamless integration of fog and cloud resources.} The system can handle the environmental heterogeneity and provide essential functions for smooth task executions. 

\textbf{End-to-end video analytics support.} The system should provide end-to-end support, including video decoding and encoding, frame processing, model inference, model tuning, and so on, ensuring users can run the pipeline with ease. 

\textbf{Low bandwidth, high accuracy.} Saving bandwidth usage while maintaining a high accuracy is the essential requirement of our system design.

\textbf{Human-in-the-Loop (HITL).}  VPaaS should incorporate human insights into the system to address the data drift issue and continually improve the model performance.

\subsection{Architecture Overview}

\begin{figure}
    \centering
    \includegraphics[width=\linewidth]{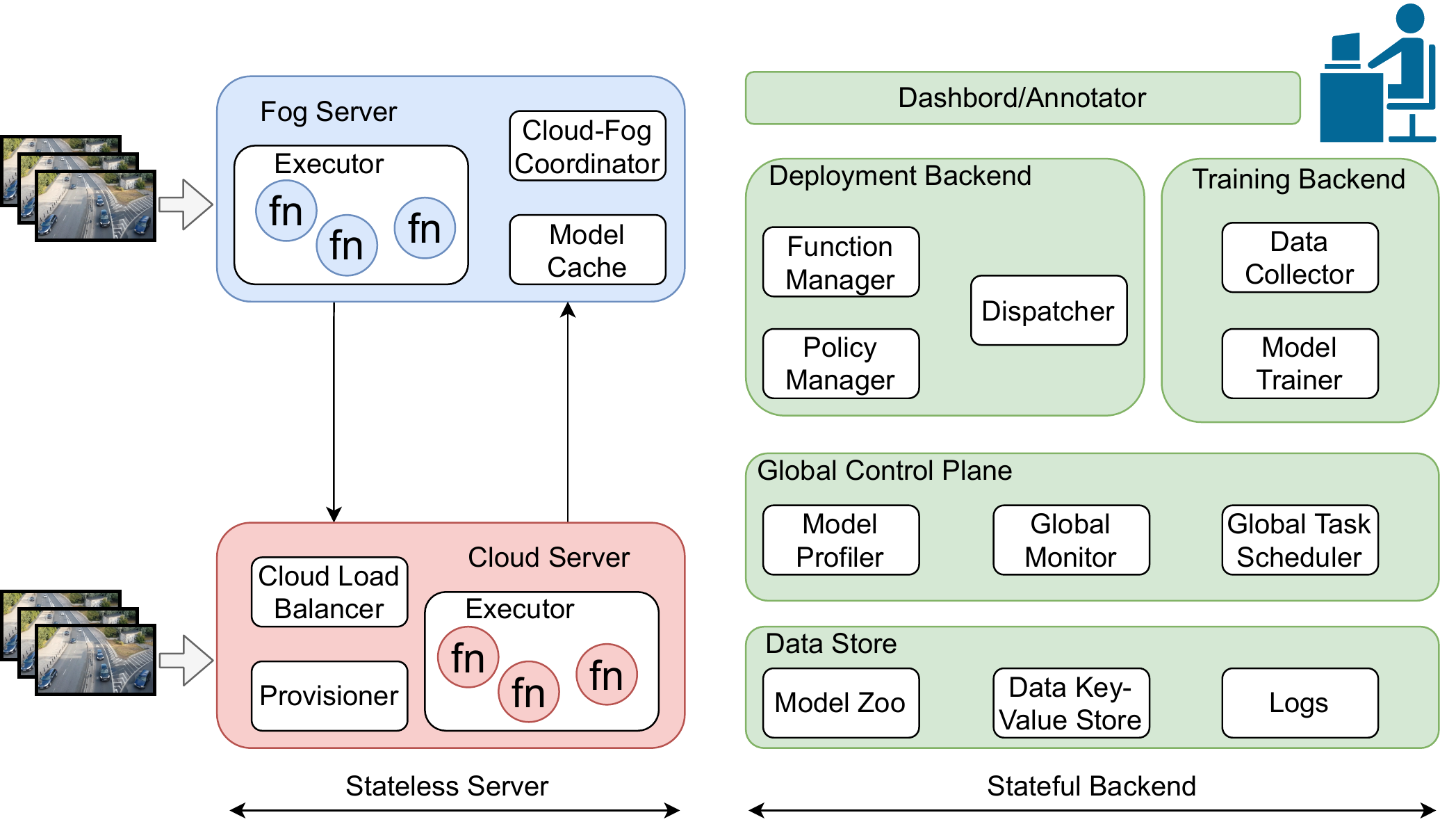}
    \caption{System architecture of VPaaS.}
    \label{fig:system_overview}
\end{figure}


Figure \ref{fig:system_overview} depicts our serverless cloud-fog platform architecture. It consists of 1) a stateless server to execute a video analytics pipeline across cloud and fog, and 2) a stateful backend to provide all essential functions to manage the whole system. Firstly, a serverless cloud-fog server is developed to serve a DNN-based video analytics pipeline. The cloud server provides an executor to run models as well as the other video processing functions. It also includes the provisioner and load balancer to provide highly available and scalable service. The fog server contains a low-latency function executor, a model cache, and a cloud-fog coordinator. The cloud-fog coordinator has the responsibility to perform the scheduling between cloud and fog with a pre-specified policy. Secondly, the stateful backend provides an interface and many functions to involve humans in the system. It allows users to register video analytics functions (e.g., newly trained ML models) and scheduling policies (e.g., ensemble) to the system. It then provides many administrative functions such as function dispatcher and model profiler to improve its usability. To continuously improve the system, the data labeled by humans will be collected, and a model trainer is used to automate the model tuning. A data storage is implemented to store models, data, and system logs. We will detail our cloud-fog coordinator and HITL design in Section \ref{sec:fog_coordinator} and Section \ref{sec:human_in_the_loop}.

\subsection{Serverless Cloud-Fog Computing}
\label{sec:serverless_cloud_fog_server}

The serverless cloud-fog ML server performs video analytics tasks, including video decoding and encoding, data pre-processing and post-processing, and ML model inference. It also provides essential functions to the stateful backend to update models (via incremental learning) and manage the system resources. 

The cloud ML server provides a runtime to execute computationally demanding tasks such as running accurate models (e.g., FasterRCNN101), training models, etc. It encapsulates all necessary functions, such as load balancer and resource provisioner, running at scale without the need for maintenance by the developers. Different from the current public serverless platform like AWS lambda that only supports CPU, our executor can utilize the GPU resources to speed up the execution of model inference tasks.

The fog ML server contains many useful functions, including a cloud-fog coordinator, a model cache, and an executor. The coordinator executes policies that need to get both cloud and fog resources involved. We design a new policy and use it to discuss the fog coordinator in detail in Section \ref{sec:fog_coordinator}. The model cache is to store models dispatched from the cloud, and the model in it will be updated periodically by our incremental learning. The executor, like that of the cloud ML server, will utilize hardware resources to run video analytics functions.

\subsection{Stateful Backend.} 

VPaaS provides users a stateful control backend, including many functions developed to support a serverless video inference serving. We organize these functions into four modules, including deployment backend, training backend, utility function, and data store. A dashboard with a video annotator is provided as a frontend to achieve HITL ML and facilitate service deployment and management.

Th deployment backend provides video analytics pipeline management and deployment across cloud and fog nodes. It includes (1) a function manager that provides a fine-grained housekeeping service (e.g., registration) for video processing related functions (as illustrated in Figure \ref{fig:example_pipelines}), (2) a policy manager that allows users to register and select scheduling policies under specific scenarios, and (3) a dispatcher for deploying functions and policies to fog and clouds.

The auto-training backend contains two useful functions: the data collector and model trainer, to automatically improve the quality of the deployed models. The data collector manages the data labeled by humans during the model inference, while the model trainer tunes models with incremental learning.

The global control plane provides all necessary system operation functions to manage and schedule system resources, freeing developers from tedious administration tasks. It has a model profiler to profile ML models on underlying fog and cloud devices, a global monitor to collect system runtime performance information, and a global task scheduler to execute the dispatched policy.

\section{Cloud-Fog Coordinator Protocol}
\label{sec:fog_coordinator}

This section presents our design and implementation for the cloud-fog coordinator. We focus on the protocol design, which is complementary to existing offloading optimizations \cite{han2016mcdnn, yi2017lavea, ran2018deepdecision, hung2018videoedge} for the resource-accuracy trade-off.  The protocol should meet the following requirements (RQ):
\begin{itemize}
    \item \textbf{RQ1.} The module can utilize the best models running in clouds to maintain high accuracy.
    \item \textbf{RQ2.} Some of the tasks can be offloaded to fog nodes with no additional cloud cost.
    \item \textbf{RQ3.} The bandwidth usage should be minimized.
\end{itemize}

To achieve these goals, we first conduct several preliminary studies and obtain some key observations and insights. Based on that, we design and implement a practical protocol named \textit{High and Low Video Streaming}.

\subsection{Key Observation and Formulation}

\begin{figure}
\centering
\begin{subfigure}{0.5\columnwidth}
  \centering
  \includegraphics[width=0.9\linewidth]{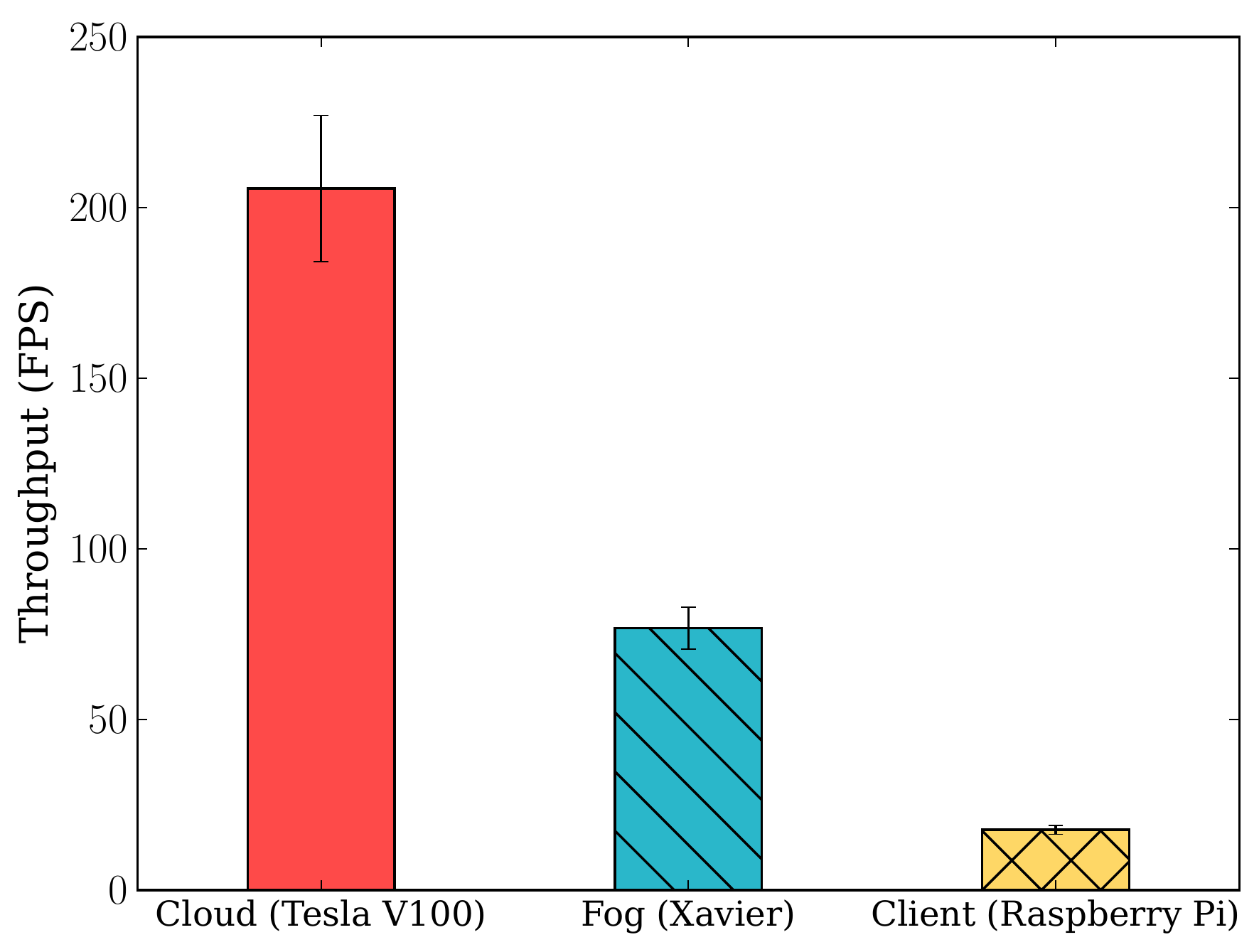}
  \caption{Quality Control}
  \label{fig:quality_control}
\end{subfigure}%
\begin{subfigure}{0.5\columnwidth}
  \centering
  \includegraphics[width=0.9\linewidth]{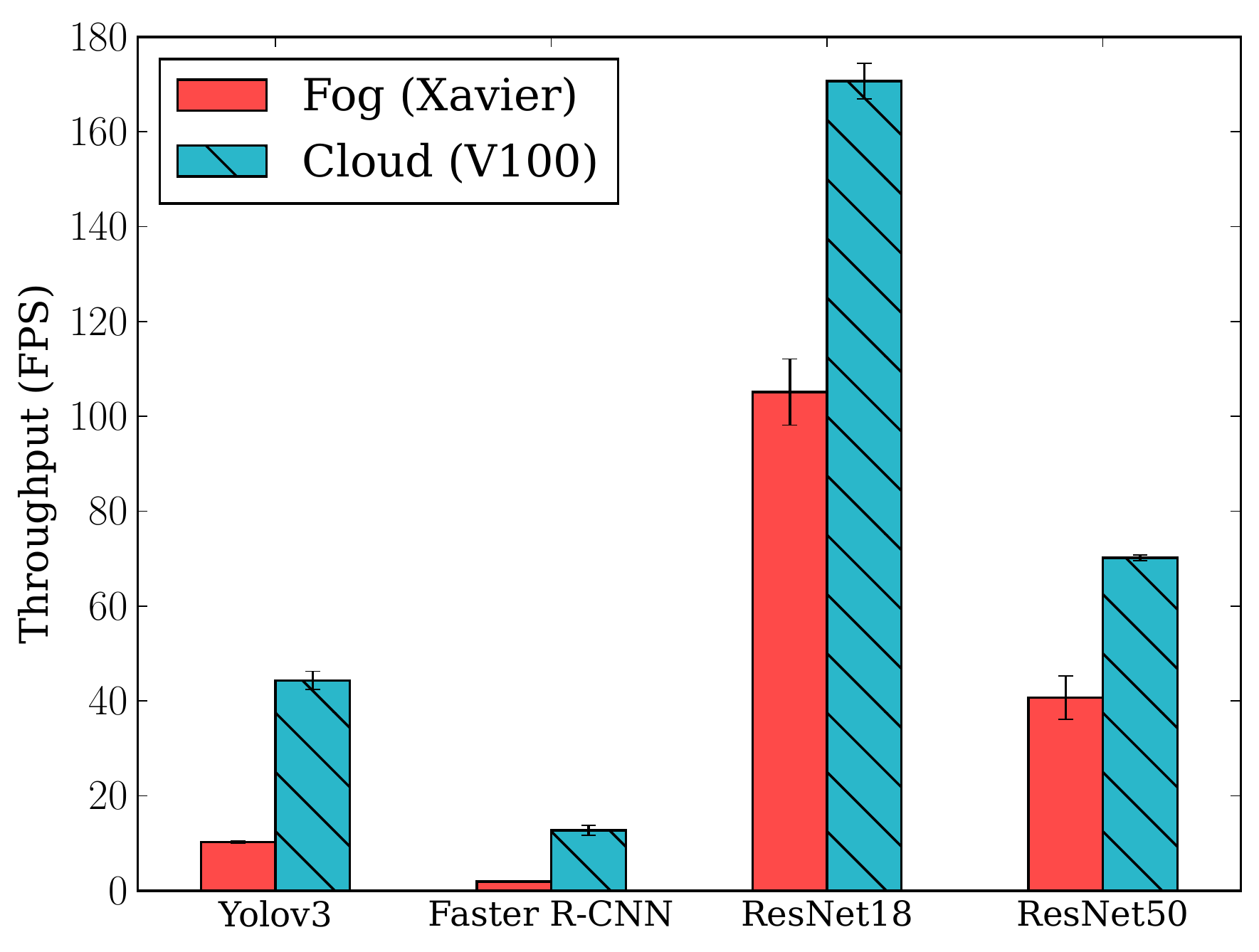}
  \caption{Model Inference}
  \label{fig:model_inference}
\end{subfigure}
    \caption{The performance of video quality control (4a) and DNN inference (4b) on different devices. Fig. \ref{fig:quality_control} shows that the computational resources for client/edge devices (Raspberry Pi 4B in the experiment) can not support real-time video decoding and re-encoding. Meanwhile, fog (NVIDIA Xavier) and cloud (V100 GPU) can efficiently perform the task. Fig. \ref{fig:model_inference} indicates that though the fog node can not support heavy object detection models very efficiently, it can more than support high-performing classification models in real-time.}
    \label{fig:edge_model_acc_speed}
\end{figure}

In our client-fog-cloud scenario, we use the following formula to summarize the dependencies of the final accuracy $a$ on design options of key system elements (e.g., choices of models and scheduling algorithms):
\begin{align}
    a = F (M_{fog}, M_{cloud}),
\end{align}
where $M_{fog}$ and $M_{cloud}$ are models running in fog and cloud, respectively. We use $F(\cdot)$ (abstractly) to describe a protocol. In this paper, we focus on finding the optimal $F^*(\cdot)$. 

Intuitively, to achieve the highest accuracy, the most straightforward protocol is to send the original quality video to the cloud where the best video analysis DNN model is running to recognize content. In this process, the bandwidth cost is incurred for transmitting video frames from the client to the cloud. This cost for a video frame is proportional to its size, which is consequently determined by the video resolution and the quantization parameter (QP) value (lower value means more details will be retained). Hence, we can denote the average video size as a function $F_v(r, q)$, where $r$ is the resolution, and $q$ is the QP value. So we can estimate the bandwidth cost for transmitting a video frame with
\begin{equation} 
B = F_v(r, q) C_B,
\end{equation}
where $C_B$ is the monetary cost for transmitting one unit of data from the client to the cloud. Developers can adjust the value according to their local data rates.

To reduce this cost, we have several design options. The first one is to avoid the transmission and only utilize the $M_{fog}$. However, this approach's inference accuracy cannot meet the requirement due to the poor performance of the compressed small models. The second approach is to rely on some filter methods like Glimpse \cite{chen2015glimpse}, only sending elaborate frames or regions to the cloud for analysis. Though this approach, in general, has a very high processing speed, they can easily miss essential frames or regions, which degrade the performance accuracy. As we will show in Section \ref{sec:experiment}, the accuracy obtained by this approach is unacceptable in our deployment scenarios.

To address these issues, we first conduct several preliminary studies. As shown in Figure \ref{fig:edge_model_acc_speed}, (\textbf{Key Observation 1}) though the fog cannot run the best object detection model very efficiently, it can support high-speed quality control and high performance classification models. In a way, this classification model is superior to detection models in terms of recognition ability. So the question becomes if we can get the regions that contain objects and only use classification models in the fog. This is precisely a chicken and egg problem: if we do not run an object detection model, how do we get the objects' locations.

\begin{figure}
\centering
\begin{subfigure}{0.5\columnwidth}
  \centering
  \includegraphics[width=0.9\linewidth]{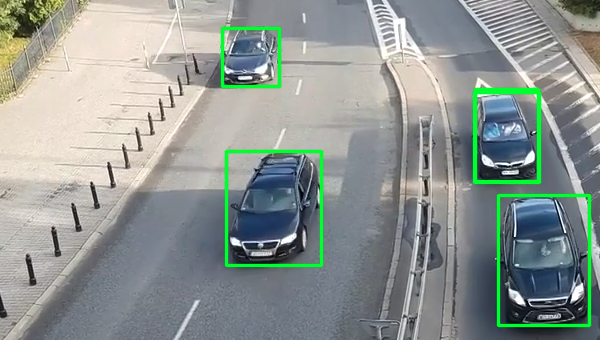}
  \caption{Ground truth}
  \label{fig:objectness_insight_groundtruth}
\end{subfigure}%
\begin{subfigure}{0.5\columnwidth}
  \centering
  \includegraphics[width=0.9\linewidth]{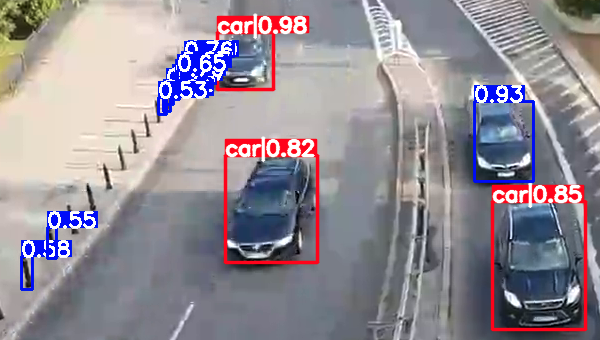}
  \caption{Results with blur videos}
  \label{fig:objectness_insight_blur}
\end{subfigure}
\caption{The output of the best cloud model with high- and low-quality videos. Fig. \ref{fig:objectness_insight_groundtruth} shows the ground truth output from the cloud model (i.e., FasterRCNN-101) for high-quality (but bandwidth consuming) videos. Fig. \ref{fig:objectness_insight_blur} illustrates that even for a very low-quality (but bandwidth efficient) video, the cloud model can output some regions that contains objects with high-confidence (red) and the locations of regions that may contain objects (blue). This observation provides us design options to save bandwidth as well as computational resources.}
\label{fig:objectness_object}
\end{figure}

To answer the question, we rely on the property of current state-of-the-art detection models (like FasterRCNN101 \cite{ren2015faster}) and continue our empirical studies. These DNNs always involve two stages. It first identifies the regions that might contain objects and then classify them into objects. We aim to utilize its location power with minimal bandwidth usage, so we adjust the resolution $r$ and QP $q$ to reduce the video size and watch its influence. As shown in Figure \ref{fig:objectness_object}, (\textbf{Key Observation 2}) even for a low quality video, the model can identify a region that possibly contains objects. It just cannot recognize the objects due to the blur video frames. A similar phenomenon has also been mentioned in \cite{du2020server}. In addition, (\textbf{Key Observation 3}) smaller video size also means lower transmission and processing time with object detection model, which leaves us much room to employ classification models on fog nodes.

\subsection{High and Low Video Streaming}

\begin{figure}
    \centering
    \includegraphics[width=0.9\linewidth]{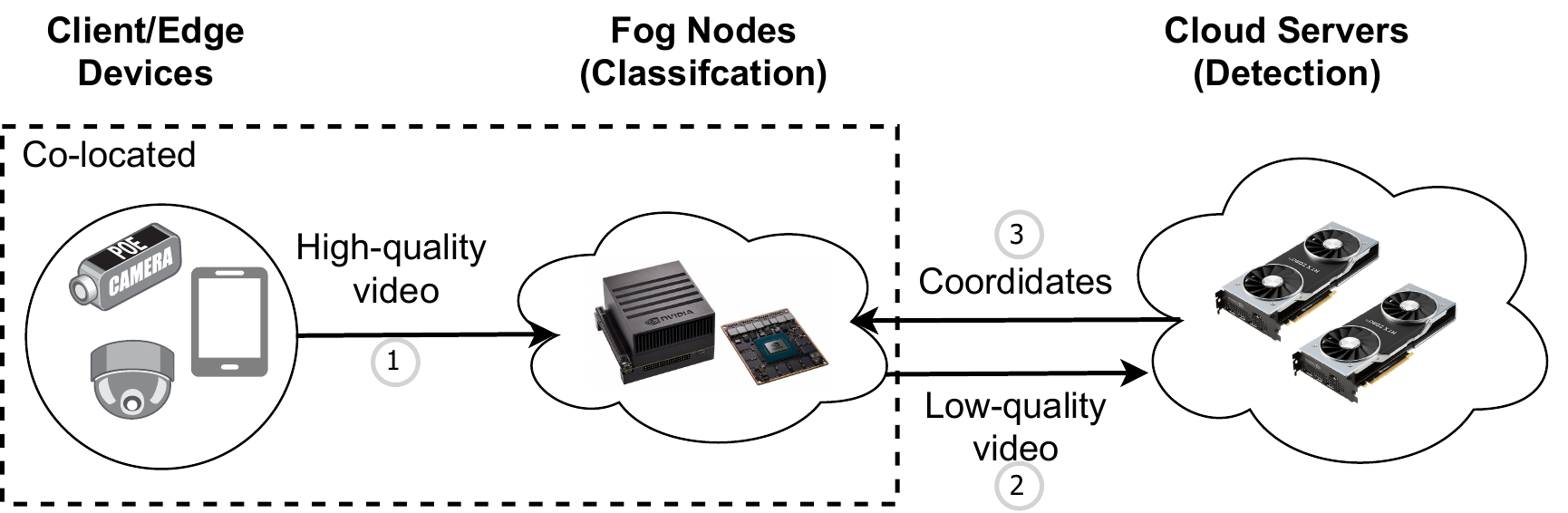}
    \caption{The overview of our cloud-fog coordinator. The client first streams high-quality videos to fog nodes, where videos will be re-encoded and sent to the cloud. The cloud runs the best model to recognize the videos. The cloud DNN models output the bounding boxes with high-confidence classification scores and only send coordinates of regions that contain uncertain objects to the fog for further processing. The design can save bandwidth while maintaining high accuracy and improve the re-encoding efficiency to reduce processing latency.}
    \label{fig:video_overview}
\end{figure}

Based on the three key observations, we now design a new protocol named \textit{high and low video streaming}. Figure \ref{fig:video_overview} depicts the whole process in our client-fog-cloud deployment scenario. First, a client equipped with a high-resolution camera will send the high-quality video to a co-located fog nodes. Since the client and fog nodes are co-located, the bandwidth cost is negligible. The video will then be re-encoded to low-quality format and sent to the cloud, where a high performing object detection model is employed to analyze the videos. The model will output the coordinates of regions that may contain objects with location confidence scores and bounding boxes with high recognition scores. We directly treat bounding boxes as labels and send them back to the fog for downstream applications. For regions that can not be recognized in a low-quality format, we apply a filter method derived from \cite{du2020server}: We first keep the regions with location confidence scores higher than a threshold $\theta_{loc}$ (for different deployment scenarios, the value may be different). We then remove the regions that have a large overlap with the above-mentioned bounding boxes. Here, we use the intersection-over-union (IoU) to indicate the overlap. An overlap value lower than a threshold $\theta_{iou}$ indicates that we can keep the region. We then remove the regions that account for more than $\theta_{back}$\% of the frame size because they are most likely to be a background.

After the filter, we will send the coordinator information of the remaining regions to fog nodes. As the information only occupies several bytes, the bandwidth usage and can be ignored. To reduce the classification overhead, we design a lightweight classification pipeline on fog by following the one-vs-all reduction rule \cite{rifkin2004defense}. The pipeline contains a feature extraction backbone network pre-trained on ImageNet dataset to learn a high-level representation from the input regions. This representation will be fed into a set of binary classifiers for classification. In doing so, we significantly reduce the computation resources needed to do the multi-class classification and maintain high accuracy (sometimes, this method can even have better performance, as illustrated in Figure \ref{fig:objectness_object}). In addition, as the video content varies within time, the number of returned coordinates will be different. To maintain high throughput and relatively low latency, we implement the well-known dynamic batching \cite{crankshaw2017clipper} and feed batched regions into the models.

\section{Human-in-the-Loop Learning}
\label{sec:human_in_the_loop}

In this section, we improve our system performance by adopting human-in-the-loop learning. We first present our observations and motivation and then describe the problem formulation and the learning process in detail.

\subsection{Observations and Motivation}

\begin{figure}
\centering
\begin{subfigure}{0.5\columnwidth}
  \centering
  \includegraphics[width=0.9\linewidth]{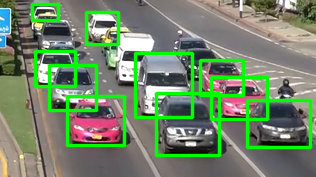}
  \caption{Detection}
  \label{fig:human_motivation_1}
\end{subfigure}%
\begin{subfigure}{0.5\columnwidth}
  \centering
  \includegraphics[width=0.9\linewidth]{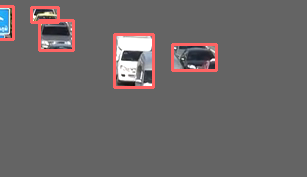}
  \caption{Classification}
  \label{fig:human_motivation_2}
\end{subfigure}
\caption{Case studies of detection and classification models. Fig. \ref{fig:human_motivation_1} shows that even for the best object detection model, some objects can not be located and recognized. Fig. \ref{fig:human_motivation_2} illustrates that once we crop the regions that can not be detected by the object detection model and feed them to a classification model, they can be classified correctly. This observation can be attributed to the misleading surrounding pixels.}
\label{fig:human_motivation}
\end{figure}

Many previous cloud-driven approaches using fixed settings rely on the predictions from the well-trained DNN running in the cloud as the ground truth \cite{jiang2018chameleon, zhang2018awstream, du2020server}. Although this can save many human efforts and speed up the system verification process, it has several drawbacks. First, as the example Figure \ref{fig:human_motivation_1} shows, many objects still can not be identified correctly even using the best model among all high-quality videos (\textbf{Key Observation 4}). Thus using the fixed setting prevents us from exploring parameter space to achieve higher performance. Second, in many cases, though an object in a completed image can not be detected and classified correctly, it can be recognized in a cropped region (\textbf{Key Observation 5}), as shown in Figure \ref{fig:human_motivation_2}. This can be attributed to the fact that surrounding pixels of an object can mislead even sophisticated models. In this case, using the prediction from the sophisticated DNN as the ground truth will result in a wrong system performance result (The system's result is right, but the \textit{ground truth} is wrong). 

Moreover, the performance of the current system relies on the fixed, pre-trained models and thus will suffer from the data drift when the distribution of online inference data diverges from the training data. Over time, the effectiveness of the entire system will deteriorate. Also, when new objects appear, the system can not handle them.

To overcome these issues, we design a module that includes human-in-the-loop and incremental learning. This particular enhancement faces two challenges. First, the module should consider the catastrophic forgetting issue, i.e., after being trained with new data, the model could perform worse on existing data). Second, the module should be lightweight, easy-to-implement, and extensible.

\subsection{Incremental Learning Process}

\begin{figure}
    \centering
    \includegraphics[width=\linewidth]{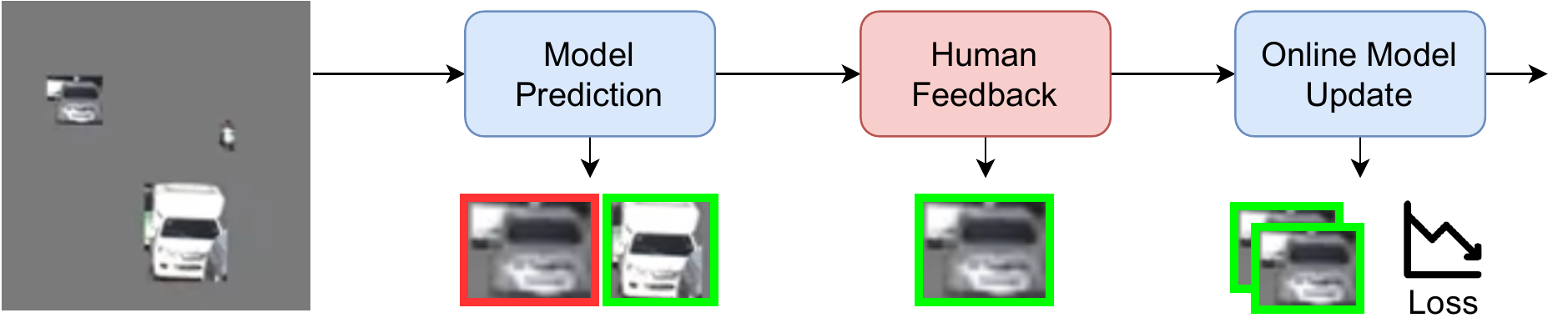}
    \caption{The human-in-the-loop process. Firstly, the cropped images will be collected together with the inference results. Secondly, annotators will check the dataset and correct the wrong results. Finally, the images with their human labels will be fed into the model for retraining.}
    \label{fig:human_in_the_loop_process}
\end{figure}

We now describe the incremental learning (IL) process, as shown in Figure \ref{fig:human_in_the_loop_process}, following our \textit{high and low video streaming protocol}. In this work, We only update the DNNs on fog servers and leave the cloud DNNs' update as the future work. We do not claim the novelty in the IL algorithm design, and the system supports easily integrating new models.

After obtaining the coordinates from the cloud, the regions that perhaps contain objects can be cropped by the system. A human operator (user) can assign a label $y_t$ to the cropped parts and a lightweight backbone network is applied to extract features vectors from them. We denote the feature vector of the $t$-th cropped image as $\textbf{x}_t$, the classifier can be given by $f(x; \Theta)$, where $\Theta$ is the set of parameters in the designed DNN. Once collecting enough images (a pre-defined value, which is decided by developers), VPaaS will update classification function $f$ in an incremental manner (to make use of human feedback on cropped images) as
\begin{align}
    f_t = f_{t-1} - \eta \partial_{f} R [f, \textbf{x}_t, y_t] \mid_{f = f_{t-1}},
\end{align}
where $R[f,\textbf{x}_t,y_t]$ is the empirical risk w.r.t. the labeled instance $(x_t,y_t )$, $\partial_{f}$ is short for $\partial / \partial f$ (gradient w.r.t. $f$), and $\eta$ is the learning rate. To achieve real-time human-in-the-loop feedback, we propose to only update weights $W$ of the last layer in the DNN. Then the objective function for updating the model is given by:
\begin{align}
\label{math:object_function}
    W = \operatorname*{arg\,min}_W \frac{1}{2} \| W - W_{t-1} \|^2_{F} + \eta l(f(\textbf{x}_t), y_t),
\end{align}
where $l(\cdot, \cdot)$ is the empirical loss and we choose it as the cross-entropy loss for classification, i.e.,
\begin{align}
    l(f(\textbf{x}_t), y_t) = y_t \text{log} f (\textbf{x}_t).
\end{align}

Here, $f(\textbf{x}_t)=\sigma(W^T \textbf{x}_t)$, where $\sigma(\cdot)$ is an activation function, and the bias term is absorbed in $W$ by simply appending the original $\tilde{\textbf{x}_t}$ with a feature 1, i.e., $W^T \textbf{x}_t = [\tilde{W}, \tilde{b}]^T [\tilde{\textbf{x}_t}, 1]$. Then the detailed formulation of (\ref{math:object_function}) can be given as follows,
\begin{align}
    \operatorname*{arg\,min}_W \frac{1}{2} \| W - W_{t-1} \|^2_{F} + \eta y_t \text{log} f (\textbf{x}_t).
\end{align}

By taking the derivative w.r.t $W$ and set it to be zero, we have:
\begin{align}
    W_t = W_{t-1} - \eta y_t \frac{1}{\sigma(W^{T}\textbf{x}_t)} \frac{\partial \sigma(W^T \textbf{x}_t)}{\partial(W)},
\end{align}
If we use $W^T_{t-1} \textbf{x}_t$ to approximate $W^T_t \textbf{x}_t$ and choose the activation function to be ReLU, then
\begin{align}
    W_t = \begin{cases}
    W_{t-1} - \eta y_t \frac{1}{\sigma(W^{T}_{t-1}\textbf{x}_t)}\textbf{x}_t, & W^T_{t-1} \textbf{x}_t > 0;\\
    W_{t-1}, & W^T_{t-1} \textbf{x}_t \leq 0.
    \end{cases}
\end{align}

When the human labor budget is exhausted after $\tau$ steps of update, we obtain a set of classifiers, i.e., $\{W_t \}^\tau_{t=1}$. These classifiers can be weighted combined to improve the performance in the further prediction. Let $\textbf{z}_i=[f(\textbf{x}_i; W_1), \cdots, f(\textbf{x}_i ; W_\tau)]$, then the weight $\omega$ can be learned by solving a regularized optimization problem, i.e.,
\begin{align}
    \operatorname*{arg\,min}_\omega \frac{1}{2} \| \omega^T \textbf{z}_i - y_i\|^2_F + v\|\omega\|^2_2,
\end{align}
where the labeled data $(\textbf{x}_i, y_i)$ obtained in the incremental learning stage is reused.

\begin{figure*}
\centering
\begin{subfigure}{0.65\columnwidth}
  \centering
  \includegraphics[width=1.0\linewidth]{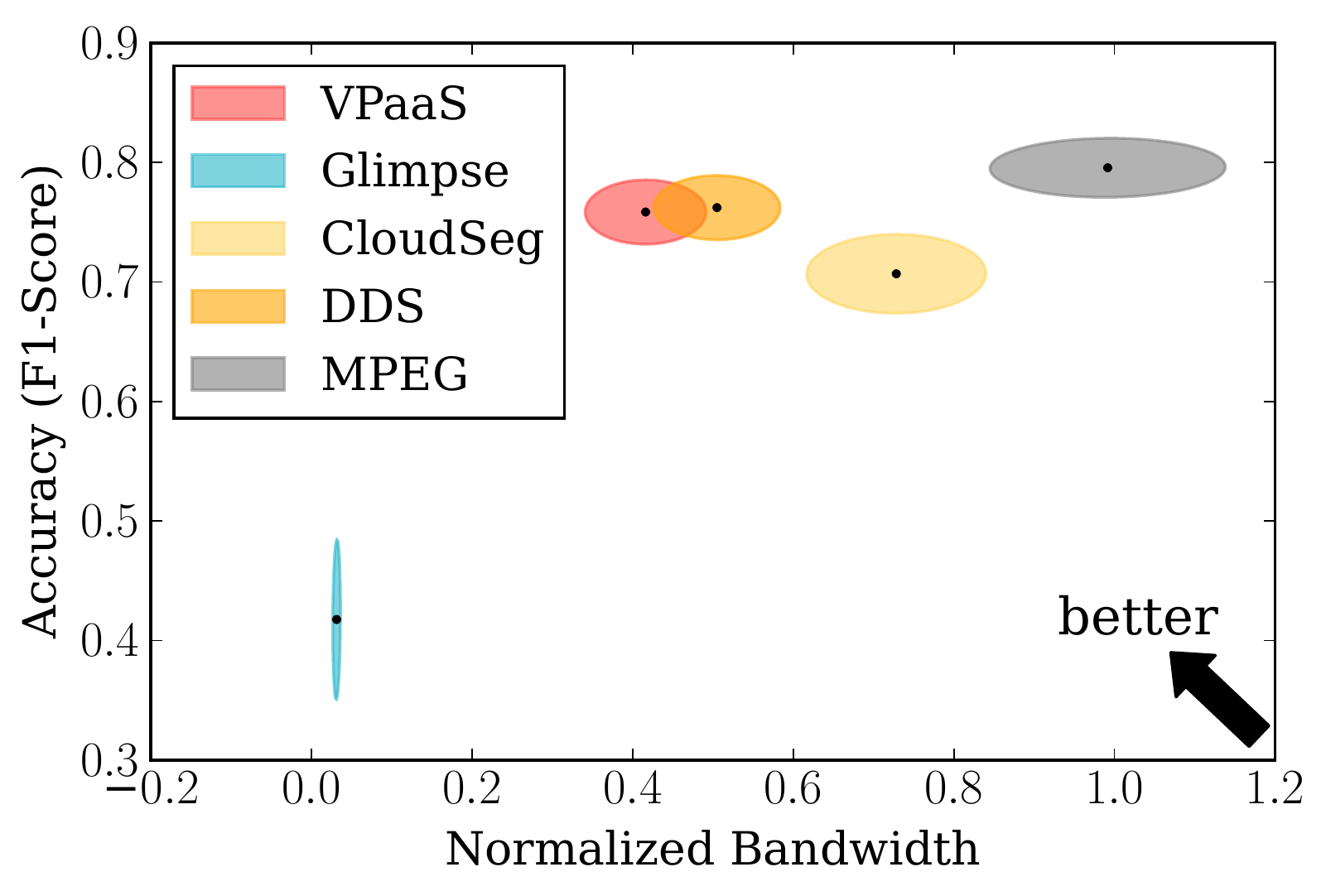}
  \caption{DashCam}
  \label{fig:dashcam_bandwidth_overall}
\end{subfigure}%
\begin{subfigure}{0.65\columnwidth}
  \centering
  \includegraphics[width=1.0\linewidth]{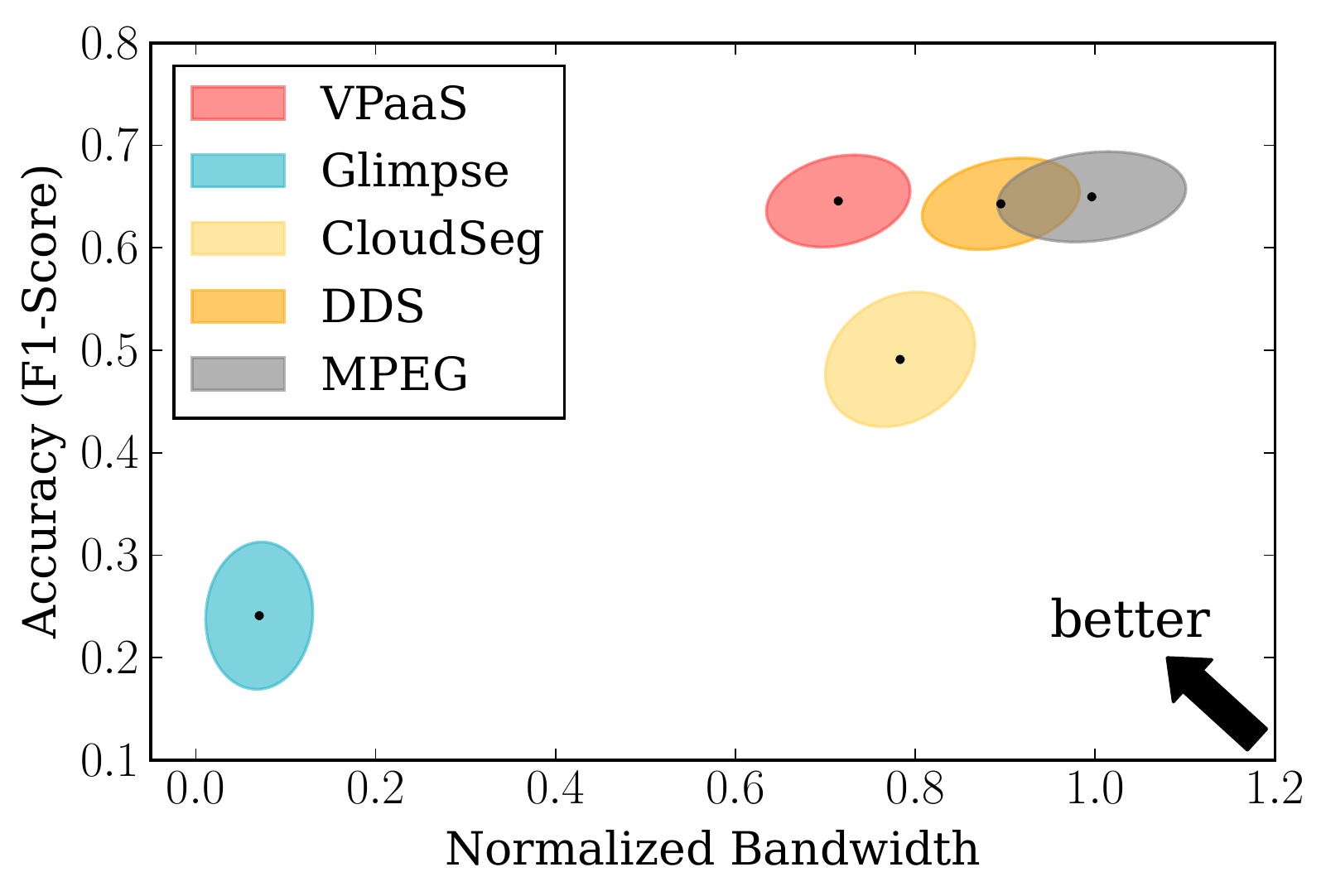}
  \caption{Drone}
  \label{fig:drone_bandwidth_overall}
\end{subfigure}%
\begin{subfigure}{0.65\columnwidth}
  \centering
  \includegraphics[width=1.0\linewidth]{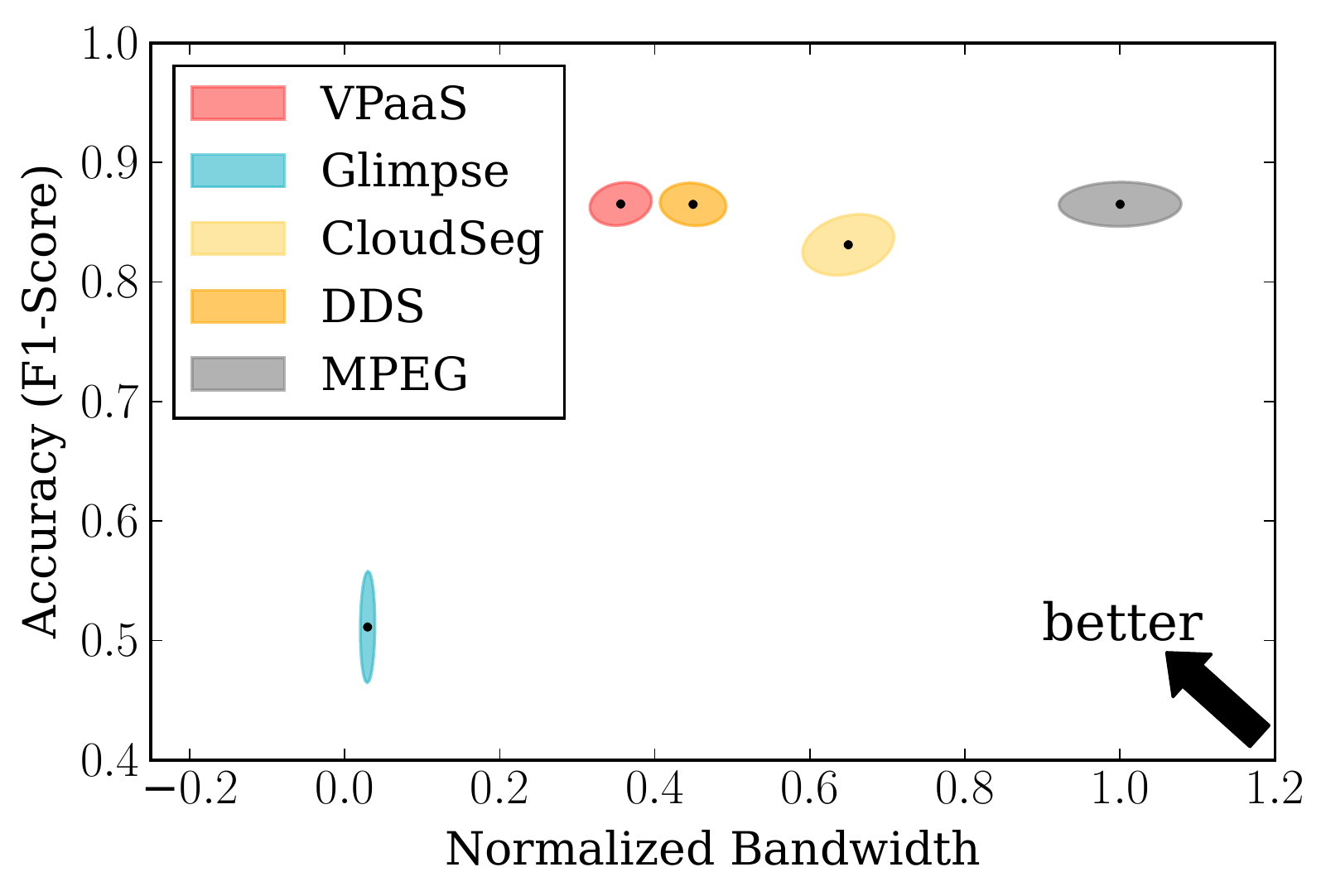}
  \caption{Traffic}
  \label{fig:traffic_bandwidth_overall}
\end{subfigure}
\caption{The normalized bandwidth usage of different systems for three video datasets. Compared to cloud-driven methods, VPaaS achieves the lowest bandwidth usage while maintaining a higher or comparable accuracy. The client-driven methods have lower bandwidth usage, but their accuracy drops drastically. The MPEG denotes using original videos to do inference.}
\label{fig:overall_bandwidth}
\end{figure*}

\section{Evaluation Results}
\label{sec:experiment}

In this section, we first introduce the system implementation details and experiment setup. We then evaluate VPaaS and present the evaluation results as well as the insights gained from them.

\subsection{Experiment Settings}

\textbf{Implementation detail.} We build our system atop CloudBurst \cite{sreekanti2020cloudburst}, an open-source, serverless platform that can be deployed to a private cluster. We extend the system to support running on cloud servers equipped with GPUs and fog devices with GPU computation cores. The communication between cloud and fog is supported by gRPC. We prototype our cloud-fog coordinator and human-in-the-loop module using Python.

We also implement a set of Python APIs to provide a fine-grained control and customization for users to perform wide-range of DNN-based video analysis tasks, such as 1) video encoding and decoding with support for many formats, 2) video pre-processing including resizing, batching, etc., 3) model inference and 4) video-post processing. We use OpenCV 4.5.0 to read videos and FFmpeg 4.3.1 to adjust the video quality. Since the DNN models are the cores for our system, we also implement a model zoo using MongoDB. We use PyTorch 1.4.0 to re-train and fine-tune our DNN models for improving them with our human-in-the-loop process.

\textbf{Experimental testbed.} We deploy our VPaaS on a real-world cloud-fog-client testbed, as shown in Figure \ref{fig:real_system}. The cloud side is hosted on servers equipped with 4 NVIDIA V100 GPUs and Intel Core i9-9940X CPU. The fog server is NVIDIA AGX Xavier \cite{nvidiaxavier}) with a 512-core Volta GPU and 8-core ARM CPU. The client is a Raspberry Pi 4B with 4GB RAM and a 1080P video camera. Following the existing video system settings \cite{zeng2020distream}, we set up a switch to build a local network for connecting the fog and clients. The bandwidth network between them is 10Gbps. Both the fog and client are connected to cloud servers through WAN (wide area network). 

\textbf{Evaluation metrics.} An ideal video analytics system should consider the following metrics, bandwidth usage, accuracy, latency, and cloud cost. 

\textit{Bandwidth Usage.} The metric evaluates the network resources used for video transmission. We calculate the bandwidth usage with $b = \frac{\sum_{0}^{n}v_i}{t}$, where $v_i$ is a video chunk with a specific quality of a video and $n$ is the total number of video chunks within a duration $t$. We normalize the usage of both our VPaaS and baselines against that of original videos without quality control.

\textit{Accuracy.} We use the F1 score (which denotes the harmonic mean of precision and recall) as the accuracy. For a public dataset without ground truth labeled by humans, we follow previous settings \cite{jiang2018chameleon, zhang2018awstream, du2020server} to run a FasterRCNN101 model and use the output from it as the label. We compare the output from a system with the label to get the true positive, false positive, and false negative, respectively, for calculating the F1 score.

\textit{Cloud Cost.} Estimating cloud cost is critical for real-world system deployment, as some scenarios require a strict cloud budget. We do not consider the costs of fog devices in our evaluations, as they are amortized to zero over the continuous frame processing. In this paper, we take the serverless billing method, which allows users to pay for the total number of requests from the public cloud (e.g., AWS). Therefore, we define the cloud cost as $c_F = p_{F}n^*$, where $p_{F}$ is the cost per frame and $n^*$ is the frame processed by the cloud.

\textit{Latency.} We measure the end-to-end latency by following the \textit{freshness} definition from \cite{du2020server} and \cite{zhang2018awstream}. It is a duration between when an object first appears in the video frame obtained by the client and when it is localized and classified in either fog or cloud. The duration consists of quality control time, transmission time, and content analytics time.

\textbf{Compared methods.} We compare our methods with three state-of-the-art methods, including one client-driven methods - Glimpse \cite{chen2015glimpse} and two cloud-driven methods - DDS \cite{du2020server} and CloudSeg \cite{234849}. Glimpse computes the pixel-level frame differences to filter out frames and runs an object tracking model to do the location and recognition. Compared to its original version, our implementation uses a more advanced tracking model from OpenCV and hence will have better accuracy. CloudSeg first sends low-quality videos from the client to the cloud and uses a pre-trained super-resolution model \cite{ahn2018fast} from its official implementation to recover the videos. Then the system will call an object detection model to analyze video content. DDS also sends the low-quality video to the cloud and then re-sends regions that may contain objects in high-quality for further process. To ensure a fair comparison, we use the same pre-trained object detection models, FasterRCNN101, in the cloud for all of these methods.

\textbf{Dataset.} We use real-world video datasets to evaluate the systems. The data covers a variety of scenarios, including traffic monitor, parking management, and video surveillance. Their details are summarized in Table \ref{tab:video_dataset} and the links to their repositories can be found in \cite{videolink}.

\begin{table}
\centering
\caption{The specifications of the video datasets used in the evaluation. Specifically, we evaluate our system with three datasets: dashcam, traffic and drone. Each of them contains a large number of video clips with a variety of scenarios.}
\label{tab:video_dataset}
\begin{tabular}{cccc}
\hline
Dataset & \# Videos & \# Total Objects & Total Video Length \\ \hline
DashCam & 3        & 46097           & 840s               \\ \hline
Drone   & 16       & 54153           & 221s               \\ \hline
Traffic & 6        & 69512           & 1547s              \\ \hline
\end{tabular}
\end{table}

\subsection{Macro Benchmarking}
\label{sec:macro_benchmarking}

We start by comparing the overall performance of VPaaS to that of other baselines on all videos.  For all of the videos under test, we follow the frame-skip setting from \cite{du2020server}, extracting one frame (called keyframes) every 15 frames. Once we accumulate 15 keyframes, we will pack them into a video chunk and then send them to the cloud. For both VPaaS and DDS, the first round QP and resolution scale (RS) are 36 and 0.8, respectively, while the second round QP and RS are 26 and 0.8, respectively. For CloudSeg, we downscale the videos with QP set to 20 and RS set to 0.35 and recover them with the upscale ratio set to 2x.


\textbf{Bandwidth and Accuracy.} Figure \ref{fig:overall_bandwidth} presents the normalized bandwidth usage and achieved F1 scores of different systems on all three datasets. We have two significant observations. First, VPaaS achieves a higher or comparable accuracy with about 21\% bandwidth saving than the closest cloud-driven system. The result indicates that by utilizing both fog computation and human-in-the-loop learning, we can further save bandwidth and improve accuracy with fewer cloud resources. Second, VPaaS consistently outperforms the client-driven methods in terms of accuracy. Though these methods consume less bandwidth, their very low accuracy stops them from being deployed to complex scenarios.

\begin{figure}
\centering
\begin{subfigure}{0.5\columnwidth}
  \centering
  \includegraphics[width=1.0\linewidth]{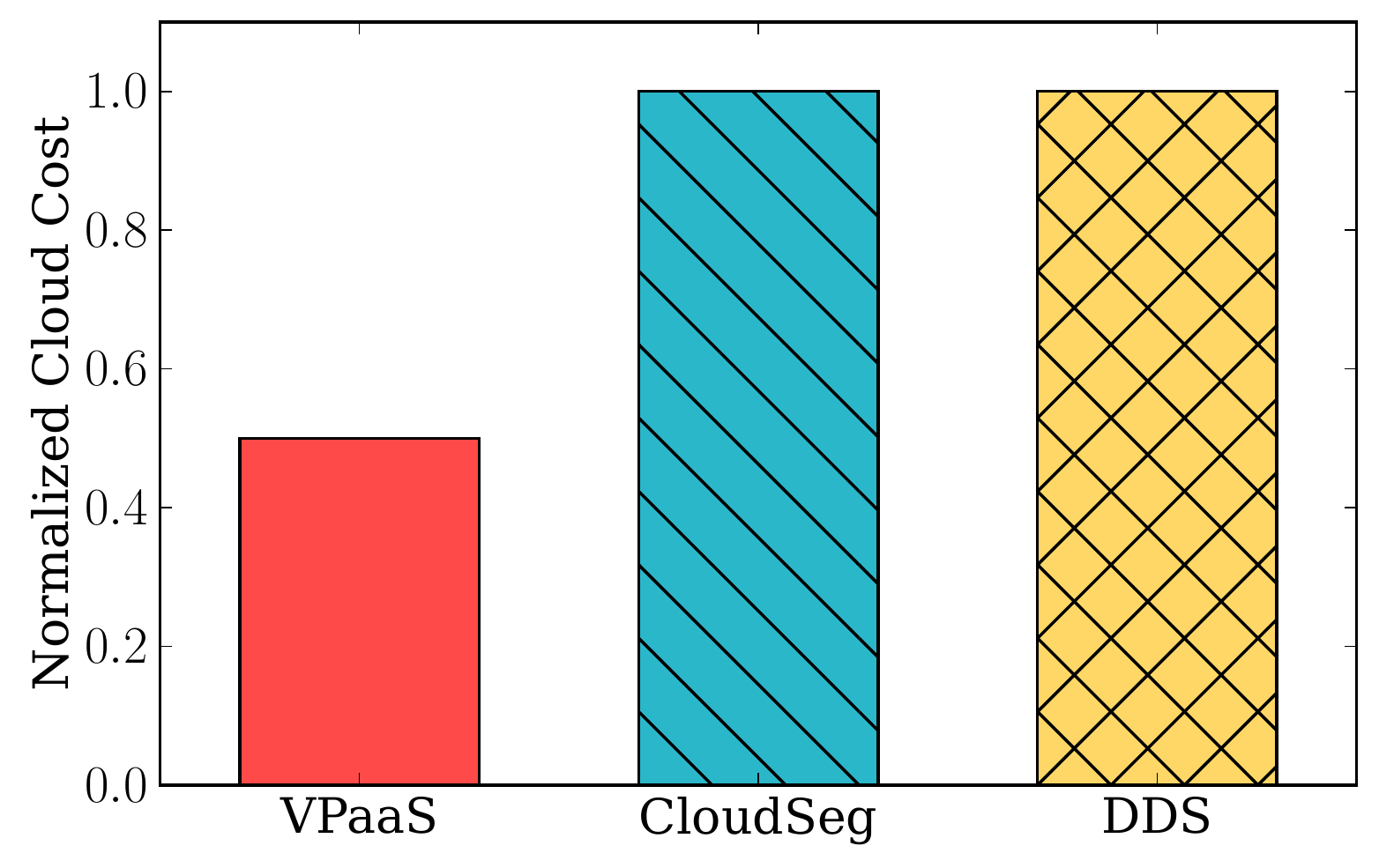}
  \caption{Cloud Cost}
  \label{fig:cloud_cost}
\end{subfigure}%
\begin{subfigure}{0.5\columnwidth}
  \centering
  \includegraphics[width=1.0\linewidth]{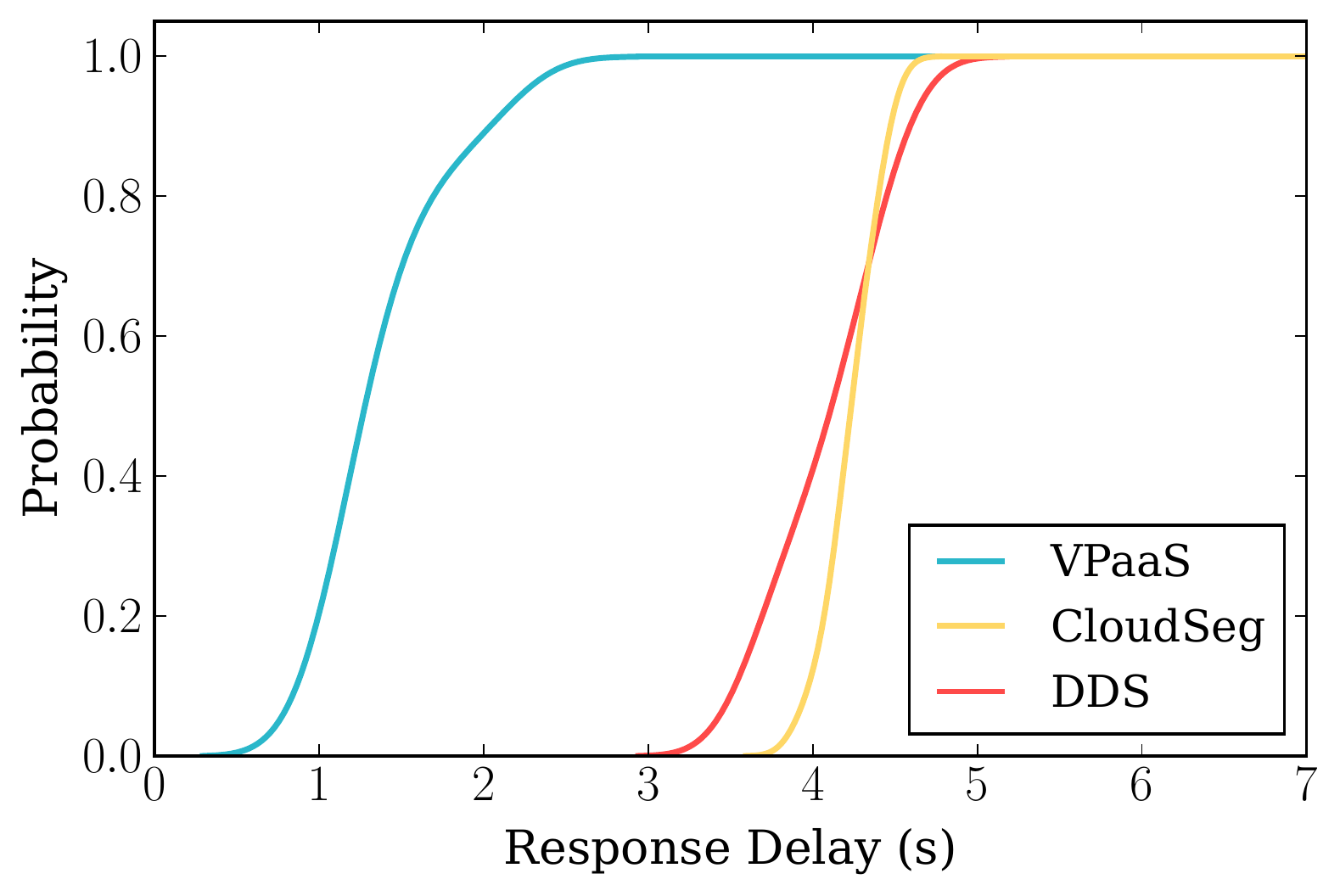}
  \caption{Latency CDF}
  \label{fig:latency_CDF}
\end{subfigure}
\caption{The normalized cloud cost and response latency for all three datasets. Compared to other cloud-driven baselines, our method do not require extra cloud resources and thus save cloud cost significantly. Meanwhile, with the help of our cloud-fog protocol design, the overall latency is reduced a large margin. We own this to the faster quality control process, near-client computation and low-quality video transmission.}
\label{fig:cloud_cost_latency_overall}
\end{figure}

\textbf{Cloud Cost.} Figure \ref{fig:cloud_cost} compares the cloud cost of VPaaS against two cloud-driven baselines. As shown, VPaaS outperforms the other two methods by a large margin. Specifically, for each frame, our system only uses the expensive object detection model running on the cloud once. Instead, CloudSeg needs an extra super-resolution model, and hence the cost is doubled compared to that incurred by our system. Also, DDS runs multiple rounds of detection on frames that are difficult to detect or classify, so it incurs more cost.


\textbf{Latency.} We report the overall latency gain in Figure \ref{fig:latency_CDF}. In general, VPaaS performs better than all the other cloud-driven methods. Specifically, VPaaS achieves about 2.5x speedup in terms of 50th (median) percentile latency than DDS and CloudSeg. Three factors contribute to the gains: moving video quality control from resource-limited clients to more capable fog nodes, reduced transmission time, and the faster classification models on fog nodes.

\subsection{Micro Benchmarking}
\label{sec:micro_benchmarking}

\begin{figure}
\centering
\begin{subfigure}{0.5\columnwidth}
  \centering
  \includegraphics[width=0.9\linewidth]{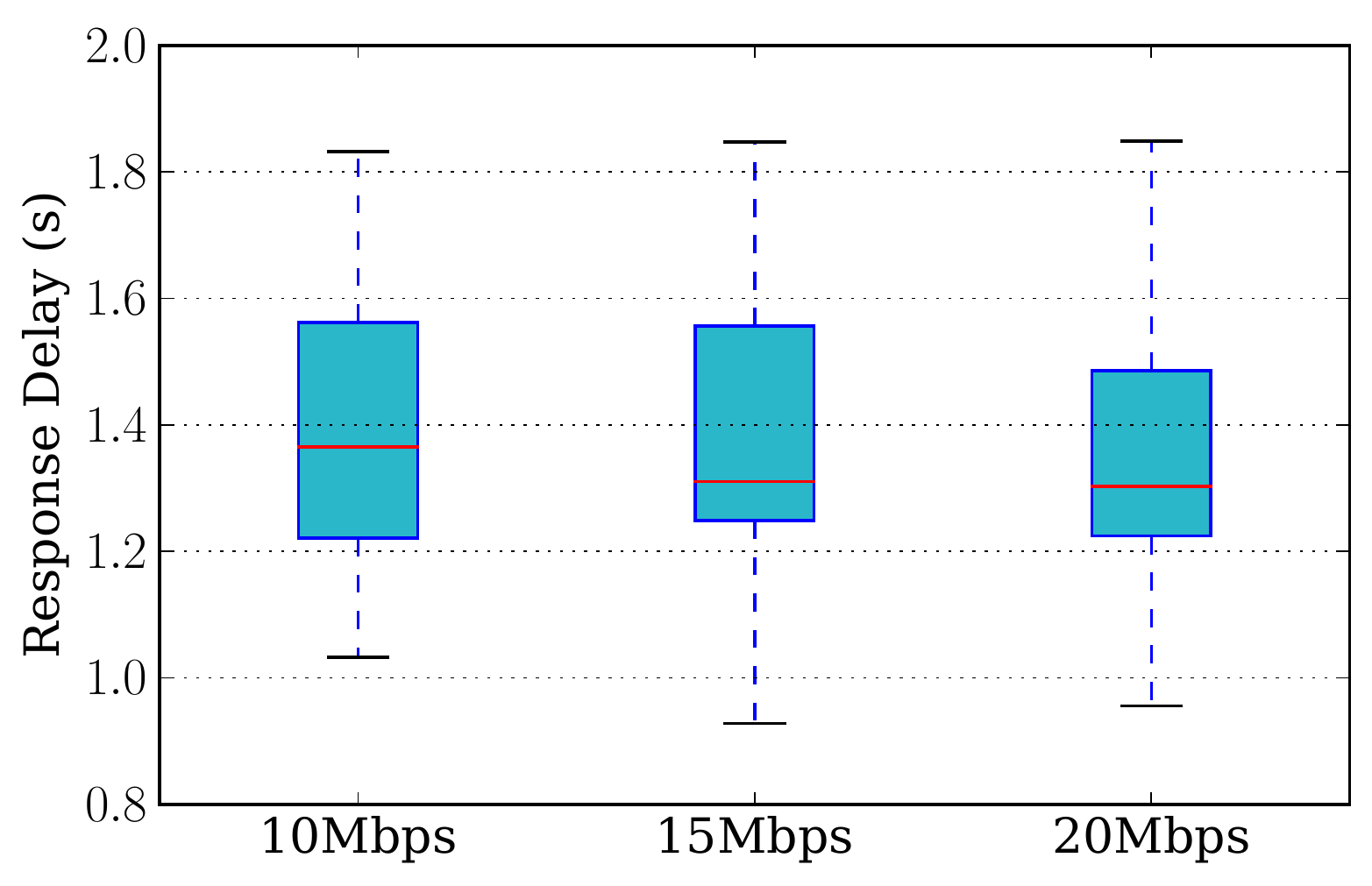}
  \caption{720P Video}
  \label{fig:bandwidth_720P}
\end{subfigure}%
\begin{subfigure}{0.5\columnwidth}
  \centering
  \includegraphics[width=0.9\linewidth]{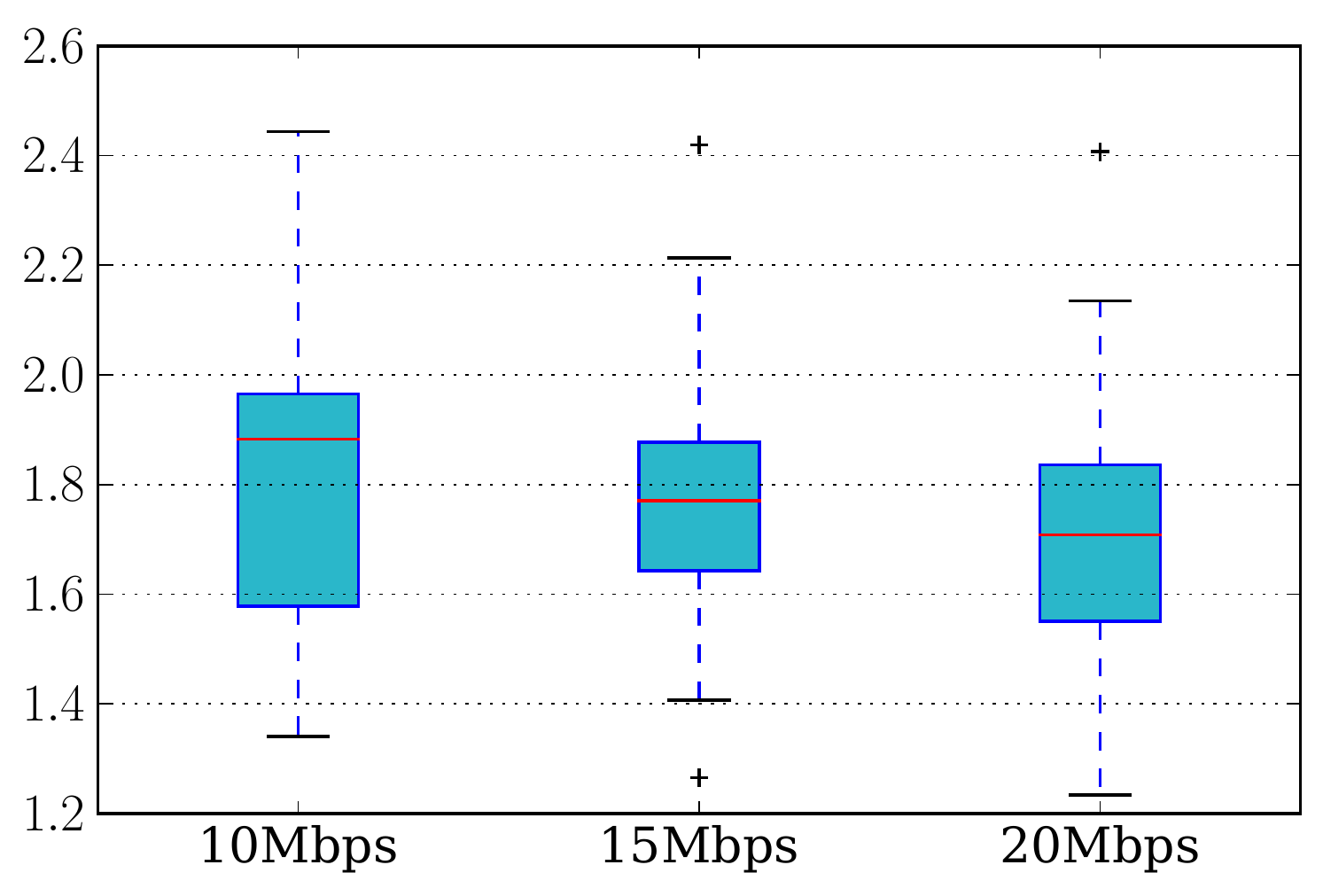}
  \caption{1080P Video}
  \label{fig:bandwidth_1080P}
\end{subfigure}
\caption{The system latency under different network bandwidth. VPaaS achieves a low latency under both low bandwidth (10Mbps) and high bandwidth (20Mbps).}
\label{fig:impact_network_bandwidth}
\end{figure}

\textbf{Impact of Network Bandwidth.} We validate VPaaS's sensitivity to network bandwidth. We test our system's response delays for a set of bandwidth [10, 15, 20] Mbps. The results shown in Figure \ref{fig:impact_network_bandwidth} demonstrate that our system can achieve a very steady latency under different network bandwidth, which indicates that our system is robust to network bandwidth fluctuations.

\begin{figure}
\centering
\includegraphics[width=0.9\linewidth]{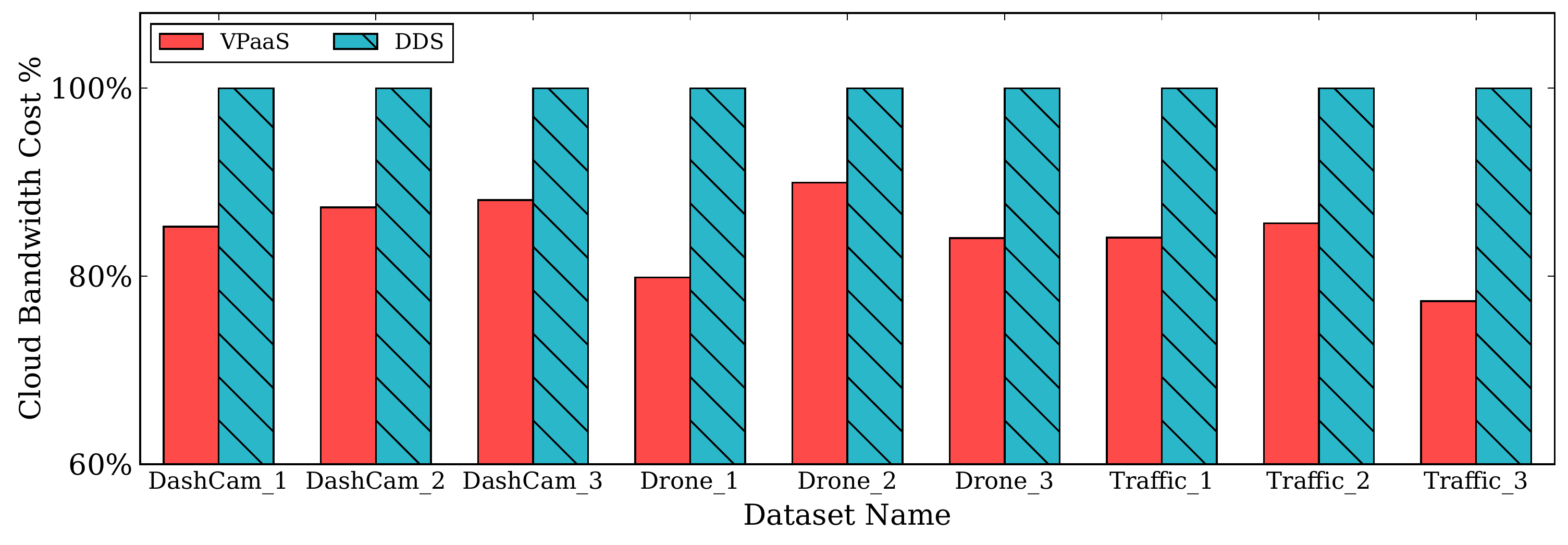}
\caption{Bandwidth usage (normalized to that of DDS) per video under three video content types. The bandwidth usage for DDS is fixed at one (100$\%$) for each content type. VPaaS outperforms the baseline in all videos, indicating that the effectiveness of our system design.}
\label{fig:impact_video_genres}
\end{figure}

\textbf{Impact of Video Content Types.} We then examine the bandwidth saving of different videos from three datasets to illustrate the impact of video content types on the performance. We first randomly select three videos from each dataset and then use the nine videos to evaluate our system and DDS (the closest work to our system). As shown in Figure \ref{fig:impact_video_genres}, our system outperforms the baseline in all video types substantially. The results prove that the performance gain mainly comes from our innovative system designs and is independent of video content.

\begin{figure}
\centering
\begin{subfigure}{0.5\columnwidth}
  \centering
  \includegraphics[width=\linewidth]{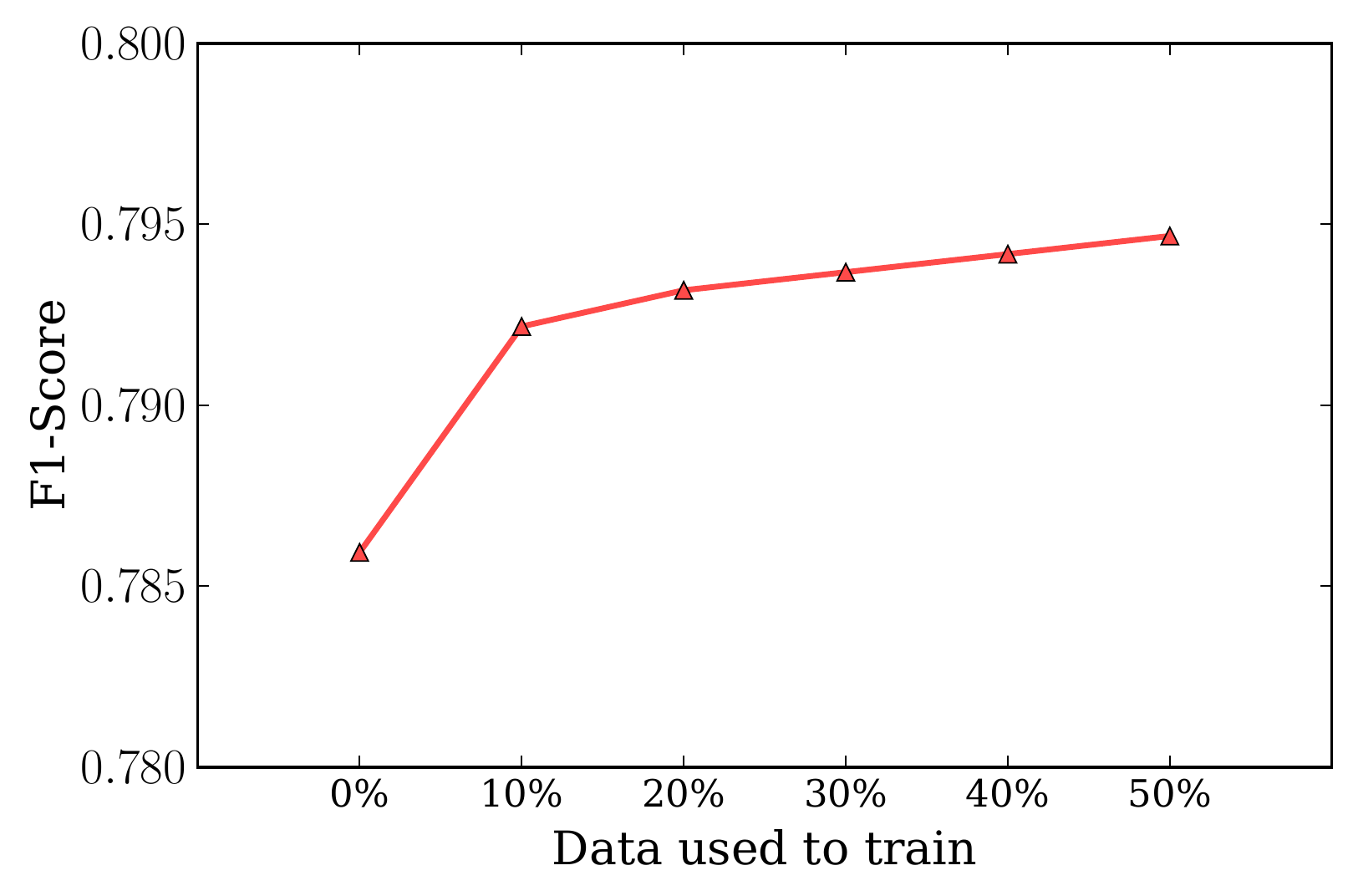}
  \caption{Human Labor Budget}
  \label{fig:labor_budget}
\end{subfigure}%
\begin{subfigure}{0.5\columnwidth}
  \centering
  \includegraphics[width=\linewidth]{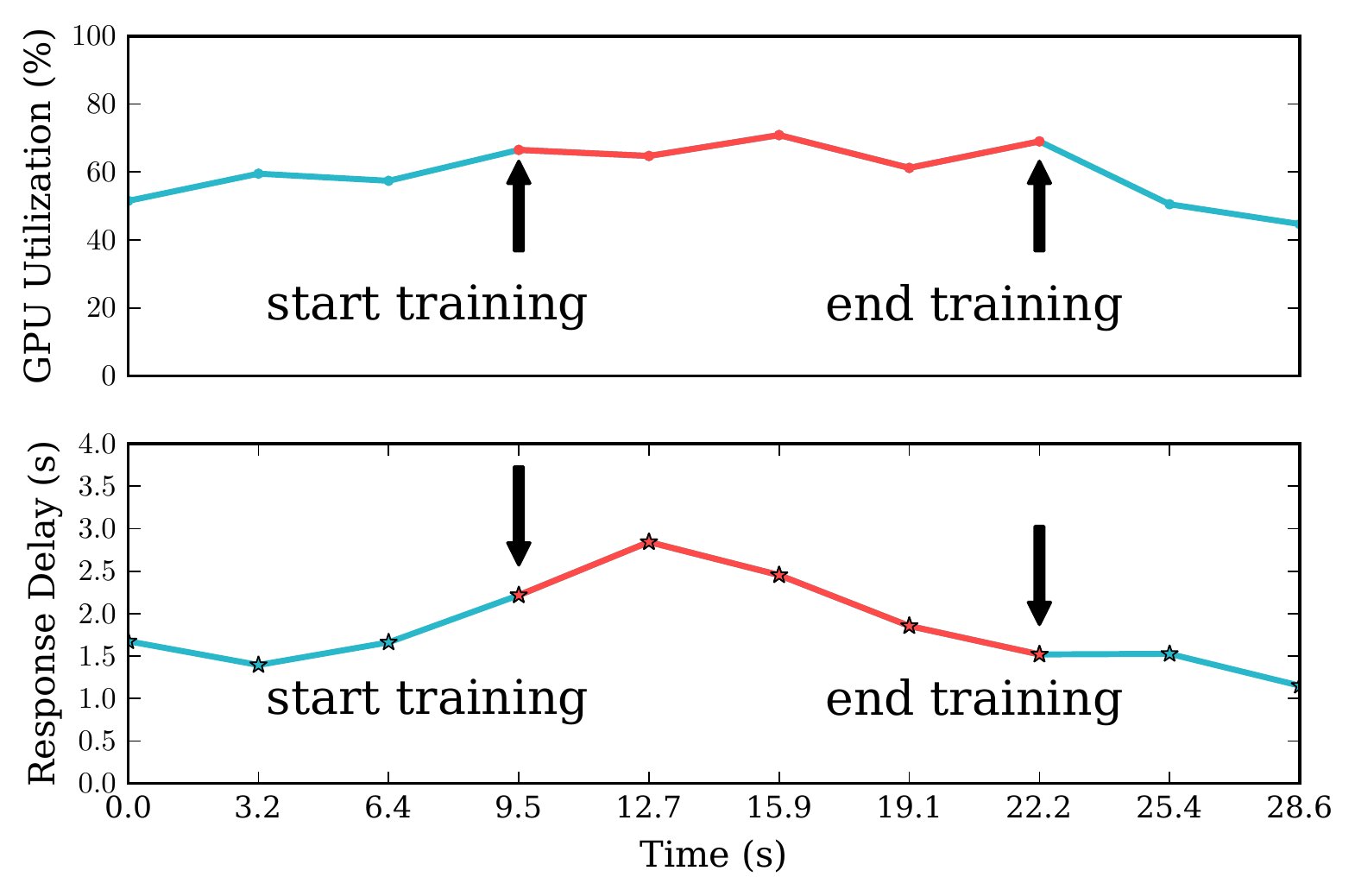}
  \caption{Overhead}
  \label{fig:overhead_human_in_the_loop}
\end{subfigure}
\caption{The impact of HITL. Fig. \ref{fig:labor_budget} shows the effect of human labor budget on accuracy, proving that the incremental learning can address the data drift issue and thus improve the performance. Fig. \ref{fig:overhead_human_in_the_loop} illustrates the training overhead. During the training, the GPU utilization (top) increases about 15\% and the latency increases about 0.5 seconds. However, the effect is considered negligible and can be further avoided by designing a better scheduler.}
\label{fig:impact_incremental_learning}
\end{figure}

\textbf{Impact of HITL Parameter.} Next, we evaluate the influence of the HITL parameter and human labor budget, which decides how much data will be used labeled in a time window. To compare the setting impact, we first divide a dataset into a training set and a test set. We use a portion of the training set for training and then gradually increase the percentage of participation in the training. The results presented in Figure \ref{fig:labor_budget} show that the HITL indeed addresses the data drift issue and improves performance. Also, as the budget increases, the effect of growth is no longer significant. We attribute this to overfitting and will explore more efficient algorithms to overcome this issue.

\textbf{HITL Overhead.} We now show the HITL overhead. During the video analytics, VPaaS triggers the auto-trainer to tune the model. It batches the data labeled by humans with batch size = 4 and feeds the batched data into the model for training. This process is executed in the same GPU for model inference to save cloud cost. Figure \ref{fig:overhead_human_in_the_loop} illustrates the overhead. During the training, the GPU utilization increases about 10\%, and the latency increases about 0.5 seconds as the result of a higher workload. Once the process is finished, the latency will quickly revert to the normal level. This finding prompts us to avoid the workload spike by starting the training process when it is relatively idle.

\begin{figure}[]
    \centering
    \includegraphics[width=\linewidth]{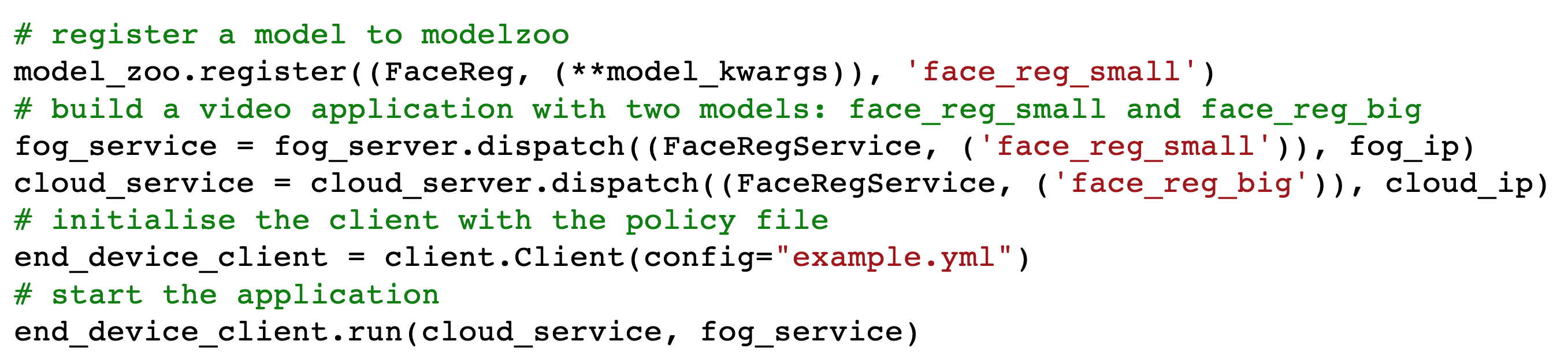}
    \caption{Example code to build a video application in VPaaS.}
    \label{fig:usability}
\end{figure}

\subsection{Case Studies}
\label{sec:case_study}

\textbf{Usability.} We show a start-to-finish process of video application development from a VPaaS user's perspective. We assume that a user has trained a face recognition model. Figure \ref{fig:usability} illustrates the user's code to build a video face recognition application across cloud and fog. Firstly, the user needs to register the model to our system, where the model will be profiled. The model with the profiling information will be stored in the cloud model zoo. Secondly, the user can dispatch the model to the fog and deploy an already-registered model to the cloud. Thirdly, the users can specify a policy to orchestrate two models for the application (e.g., monitoring the networking congestion/latency to decide whether to send videos to the cloud or process them locally). In addition, our system will automatically call the video pre- and post-processing functions to complete the pipeline.

\begin{figure}
    \centering
    \includegraphics[width=0.9\linewidth]{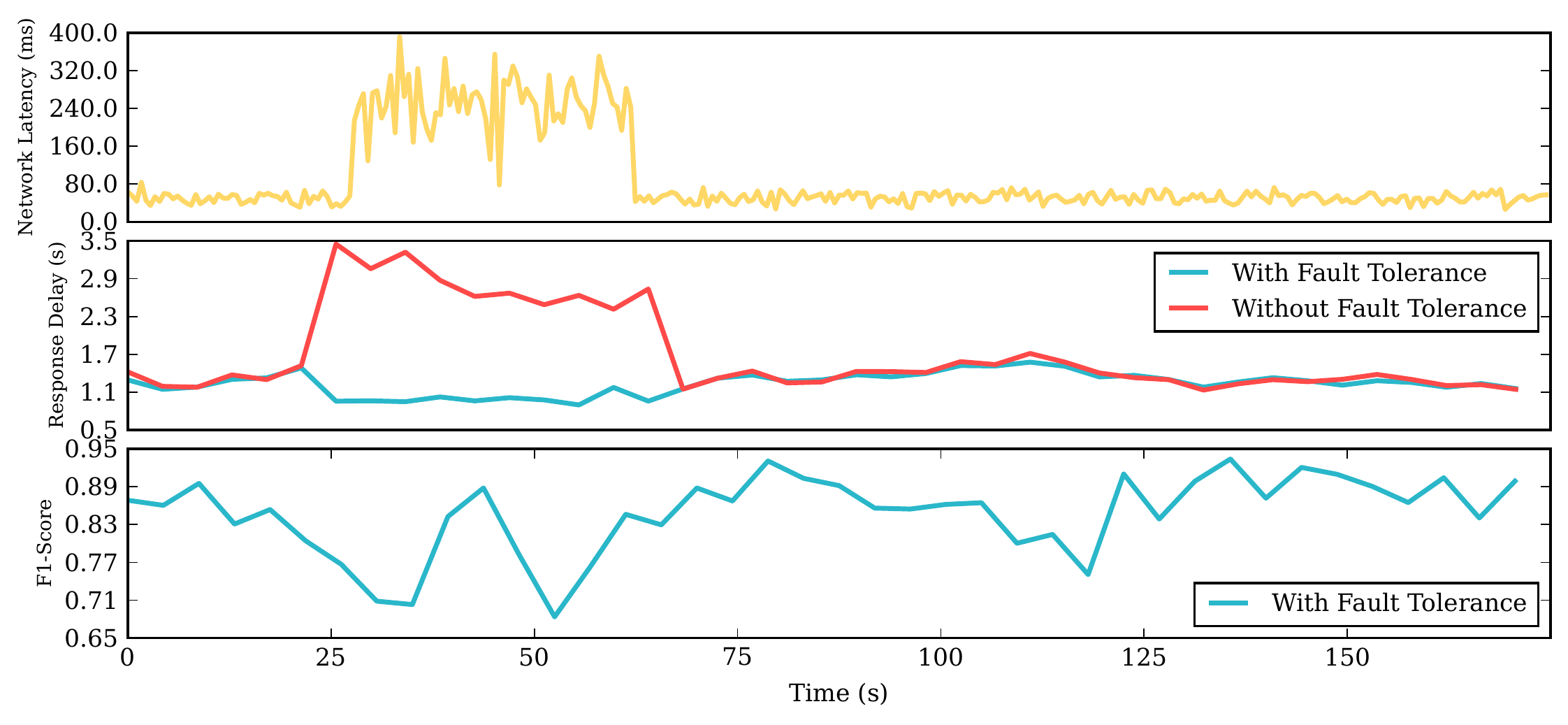}
    \caption{Fault-tolerance evaluation. VPaaS can quickly call a small backup object detection model (e.g., YOLOv3) on fog once detecting the networking disconnection issue. The bottom figure shows the fluctuations in accuracy. Though the accuracy drops but the system can still provide a low-latency service (middle).}
    \label{fig:fault-tolerance}
\vspace{-0.2in}    
\end{figure}

\textbf{Fault-tolerance.} To test the system's fault-tolerance feature, we simulate an outage scenario by shutting down the cloud server. In this situation, our fog nodes will run a backup, using a light-weight object detection model such as YOLOv3 \cite{redmon2018yolov3} to resume the recognition tasks quickly (albeit with reduced accuracy). As shown in Figure \ref{fig:fault-tolerance}, the fog node detects the disconnection issue at $t$=25s. Then it feeds the video chunks that are already cached in fog nodes to the YOLOv3 model deployed at the fog to continue object detection tasks. Though the accuracy decreases, our system can maintain the service continuation until the recovery of the cloud server.

\begin{figure}
    \centering
    \includegraphics[width=0.9\linewidth]{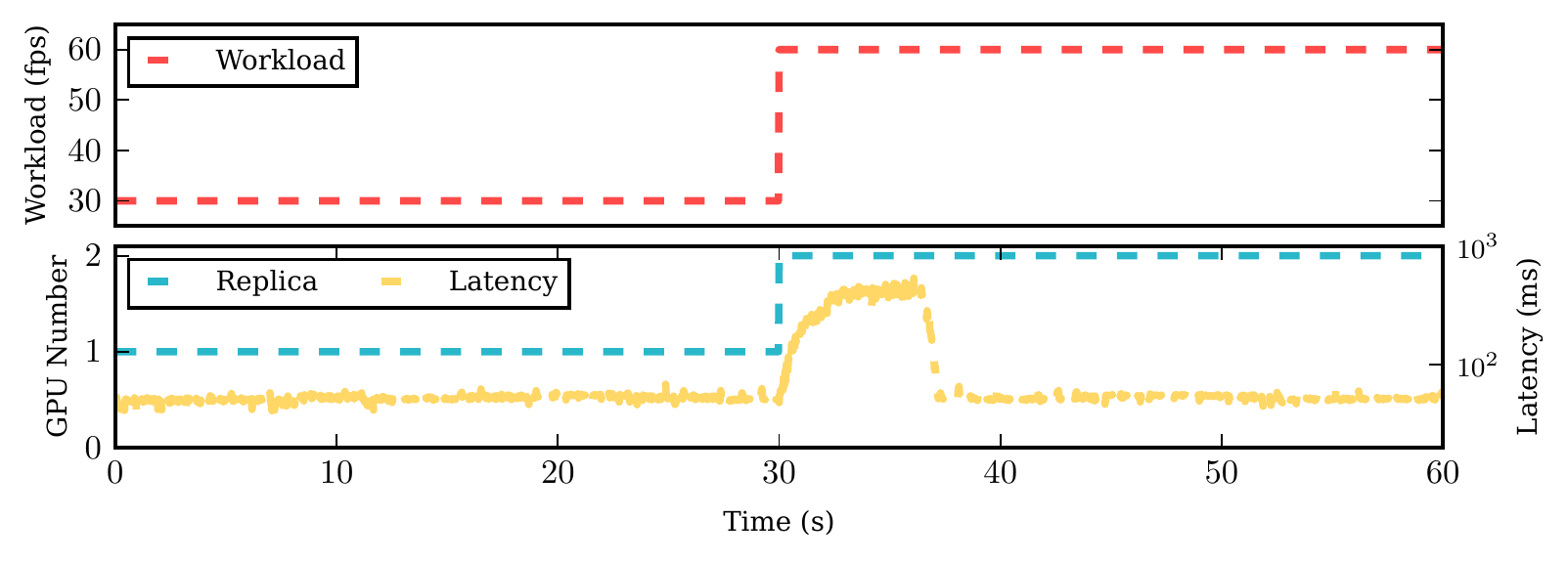}
    \caption{Scalability evaluation. Our serverless platform can scale in/out GPUs to save cost and maintain high availability when experiencing dynamic workload. In this case, we simulate a scenario where users install more fog nodes and cameras for their video applications.}
    \label{fig:scalability}
\end{figure}

\textbf{Scalability.} Our serverless system also provides the essential provision function for the dynamic workload. To test the feature, we first simulate a scenario where users install more fog nodes and cameras by increasing the number of video chunks sent simultaneously. As shown in Figure \ref{fig:scalability}, the number of GPUs used increases as the more video chunks come in, so that our system maintains a low latency even during heavy workload.

\section{Conclusion}
\label{sec:conclusion}

Efficient video analytics empower many applications ranging from smart city to warehouse management. This paper presents a serverless platform termed VPaaS to run DNN-based video analytics pipelines that take full advantage of the client-fog-cloud infrastructure's synergy. It can efficiently orchestrate both fog and cloud resources for cost-effective and high-accurate video analytics. VPaaS employs human-in-the-loop design philosophy, continuously improving model performance. The system provides a set of functions for video application development and deployment, freeing developers from tedious resource management and system administration tasks. Extensive experiments demonstrate that VPaaS consumes less bandwidth and cloud cost and has lower processing latency than state-of-the-art systems. We plan to involve network topology in the system design as future work. Also, preserving privacy in video analytics is very important, and we plan to explore this research direction in the future.





\bibliographystyle{IEEEtran}
\bibliography{fogvideo}

\end{document}